\documentclass[aps,showpacs,11pt]{revtex4}
\usepackage{dcolumn}
\usepackage{graphicx}
\usepackage{amsmath}
\usepackage{amsfonts}
\usepackage{amssymb}
\usepackage{psfrag}
\usepackage{wrapfig}
\usepackage{subfigure}
\usepackage{makeidx}
\usepackage{bm}
\usepackage{epsf}
\usepackage{hyperref}
\usepackage{color}
\usepackage{float}
\usepackage{ulem}

\def\/{\over}

\begin{document}

\title{Early dark energy and its interaction with dark matter}

\author{Bo-Yu Pu, Xiao-Dong Xu, Bin
Wang} \affiliation{IFSA Collaborative Innovation Center, Department of Physics and
Astronomy, Shanghai Jiao Tong University,
Shanghai 200240, China }
\author{Elcio Abdalla}
\affiliation{Instituto de Fisica, Universidade de
Sao Paulo, CEP 05315-970, Sao Paulo, Brazil }

\begin{abstract}
We  study a class of early dark energy models
which has substantial amount of dark energy in
the early epoch of the universe.  We examine the
impact of the early dark energy fluctuations on
the growth of structure and the CMB power spectrum in
the linear approximation. Furthermore we
investigate the influence of the interaction
between the early dark energy and the dark matter
and its effect on the structure growth and CMB. We
finally constrain the early dark energy model
parameters and the coupling between dark sectors
by confronting to different observations.
\end{abstract}

\pacs{98.80.-k}


\maketitle

\section{Introduction}
From astronomical observations, it is convincing
that our universe is undergoing accelerated
expansion. The driving force of this acceleration
is dark energy(DE), which composes roughly $70\%$
of the total energy budget of our universe.  The
physical nature of DE, together with its origin
and time evolution, is one of the most enigmatic
puzzles in modern cosmology. The simplest
explanation of DE is the cosmological constant
with equation of state (EoS) $w=-1$. Although the
cosmological constant fits well to current
observational data, it suffers serious
theoretical problems. One is the cosmological
constant problem, the fact that the quantum field
theory prediction for the value of $\Lambda$ is
about hundred orders of magnitude larger than
the observation \cite{coincidence}. Another
problem, more closely related to the cosmological
evolution itself, is the coincidence problem,
namely why being a constant, the $\Lambda$ value
becomes important for the evolution of the
universe just at the present moment \cite{coin}.
Besides the cosmological constant, there are
other alternative explanations of DE. But so
far, the focus has been on the EoS of DE and in
particular on its current value $w_0$.

It is rather the amount of DE, $\Omega_{de}$, than
the EoS, that influences the evolution of our universe.
In this spirit, an interesting sub-class of DE
models  involving a non-negligible DE
contribution at early times has been proposed.
These models are called Early Dark Energy (EDE),
and have been extensively studied recently. EDE
models can potentially alleviate the coincidence
problem. Furthermore, they can influence the
cosmic microwave background
\cite{1304.3724,Doran2001,1003.1259,astro-ph/0508132,astro-ph/0611507,1305.2209,0901.0605},
big-bang nucleosynthesis\cite{Muller} and
large-scale structure formation
\cite{Bartelmann,1004.0437,1207.1723,0809.3404,1303.0414,xx,Baldi}.
For now, it would be fair to say that there are
no strong observational constraints on the EDE
models, and it is especially difficult to
discriminate EDE models which have $w=-1$ at
present from the $\Lambda$CDM model.

In this paper, we will focus on a specific EDE
model, which is similar to that originally
introduced by Wetterich \cite{wetterich}
 and further examined in \cite{0809.3404}.
This model is characterized by a low but
non-vanishing DE density at early times with the
EoS varying with time in the form
\begin{eqnarray}
w(z)=\frac{w_{0}}{1+b\ln{(1+z)}^2}, \hspace{1cm}
b=-\frac{3w_{0}}{\ln{(\frac{1-\Omega_{de,e}}{\Omega_{de,e}})}+\ln{(\frac{1-\Omega_{m,0}}{\Omega_{m,0}})}}\label{1}
\end{eqnarray}
where $w_{0}$ and $\Omega_{de,0}=1-\Omega_{m,0}$
represent the present-day EoS and amount of DE,
respectively, while $\Omega_{de,e}$ gives the
average energy density parameter at early times. The EDE model described in (1) has been confronted to the type Ia supernova test including samples at $z>1.25$ \cite{Rahvar}\cite{yungui}.
The impact of this EDE cosmology on galaxy
properties has been studied by coupling
high-resolution numerical simulations with
semi-analytic modeling of galaxy formation and
evolution \cite{1207.1723}. The available results
highlight that such EDE model leads to important
modifications in the galaxy properties with
respect to a standard $\Lambda$CDM universe.

We use this dynamical EDE parametrization to
further discuss the influence of this specific
model on the cosmic microwave background
radiation (CMB) and compare with the $\Lambda$CDM
prediction.   For dynamical DE models, in contrast
with $\Lambda$CDM, they possess DE fluctuations.
In the linear regime, these fluctuations for
usual DE models, for example quintessence, are
usually several orders of magnitude smaller than
that of dark matter (DM), so that DE fluctuations
are usually neglected in studies of CMB and
structure formations in the linear approximation.
It would be interesting to examine the presence
of the EDE fluctuation and its impact on the DM perturbations
and CMB, and compare with the usual
assumption of nearly homogeneous EDE and
$\Lambda$CDM models. This can help to distinguish
between homogeneous and inhomogeneous EDE models
and also disclose the difference from the
$\Lambda$CDM model. Moreover we will attempt to
constrain this EDE model using current data. In \cite{0901.0605} one different type of EDE model was constrained by using observational data at high redshift including the WMAP five year data for CMB, but in their study they have not compared different effects brought by the inhomogeneous and homogeneous EDE.

It is clear that DM and DE are two main
components of our universe, which compose almost
$95\%$ of the total universe.
It is a special assumption that these two biggest components existing
independently in the universe. A more natural
understanding, in the framework of field theory, is
to consider that there is some kind of interaction
between them.  It has been shown that the
interaction between DM and DE is allowed by
astronomical observations and can help to
alleviate the coincidence problem, see for
example \cite{1311.7380, 1308.1475, 1012.3904, 1001.0097, 0710.1198}
and references therein. It would be of great
interest to extend the previous studies to the
interaction between EDE and DM. With the
non-negligible DE energy density at high
redshift, the interaction between dark sectors
will start to play the role earlier. To
investigate the influence of the interaction
between EDE and DM on the structure growth and CMB
signals is the second objective of this paper.

The outline of the paper is the following. In the
next section, we will first present the
background evolution of the EDE model and discuss
the influence of the interaction between dark
sectors on the background dynamics. And then we
will study evolutions of linear perturbations of
a system with EDE and pressureless matter and
calculate the growth of structure. We will examine the
effect of the interaction between EDE and DM on
the linear perturbations. Section III is devoted
to the study of the CMB power spectrum. In
Section IV we will present the constraint of the
EDE model from fittings to current observational
data and in the last section we will present our
conclusions.

\section{Analytical Formalism}

In this paper, we investigate the EDE model presented
in (\ref{1}), in which there is a low but non-vanishing DE density at
early times. We modified CAMB code to examine the influences of the EDE
on the background evolution, linear perturbation
and CMB power spectrum by performing analysis for two
models `EDE1' and `EDE2', which have $w_0=-0.93,
\Omega_{de,e}=2\times 10^{-4} (b=0.29,
\Omega_{m,0}=0.25)$ and $w_0=-1.07,
\Omega_{de,e}=2\times 10^{-4} (b=0.33,
\Omega_{m,0}=0.25)$, respectively.

\begin{figure}
\centering
\includegraphics[width=0.5\columnwidth]{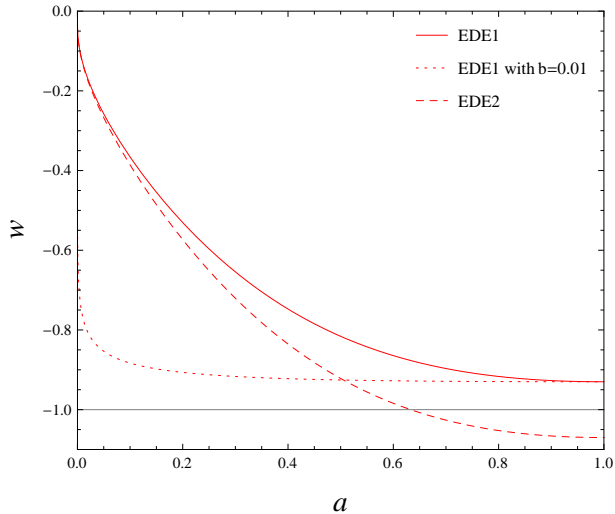}
\caption[width=1\columnwidth]{\label{w_diff_EDE}EoS of two EDE models. The dotted line refers to the EDE1 with b=0.01. }
\end{figure}

Figure \ref{w_diff_EDE} shows the evolutions of
EoS in EDE models that we examine in this work.
The amount of DE at early times is non vanishing
and EDE models approach to the cosmological
constant scenario at recent times. The EDE1 model
has EoS always above $-1$, while EDE2
EoS can cross $-1$ and stay below $-1$ at
present.

\begin{figure}[htp]
 \begin{tabular}{cc}
    \begin{minipage}[t]{3in}
    \includegraphics[width=1\columnwidth]{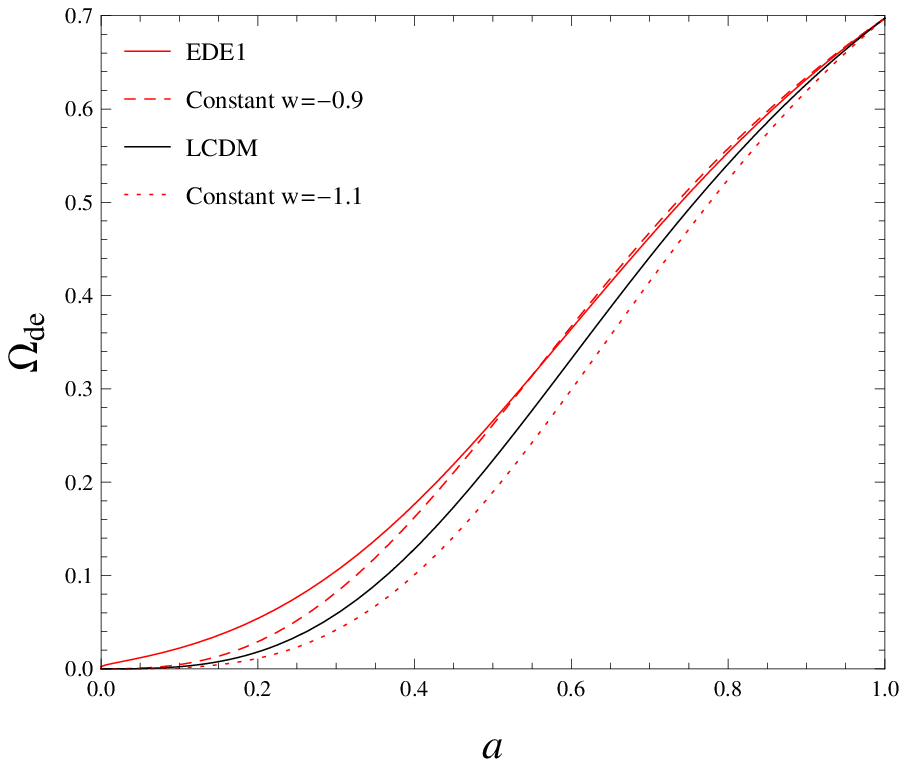}
    \end{minipage}
    \begin{minipage}[t]{3in}
    \centering
    \includegraphics[width=1\columnwidth]{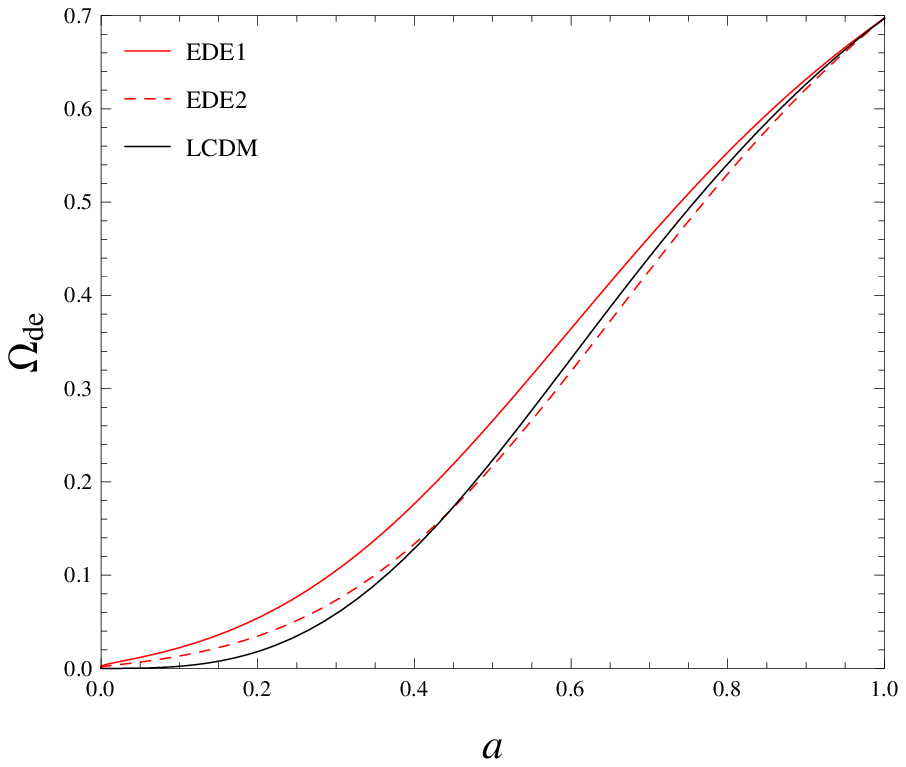}

    \end{minipage}
\end{tabular}
\caption{\label{bg_EDE_const}The evolutions of DE
fractional energy densities for different DE
models when there is no interaction between dark
sectors.  }
\end{figure}

\begin{figure}[htp]
\centering
\includegraphics[width=0.5\columnwidth]{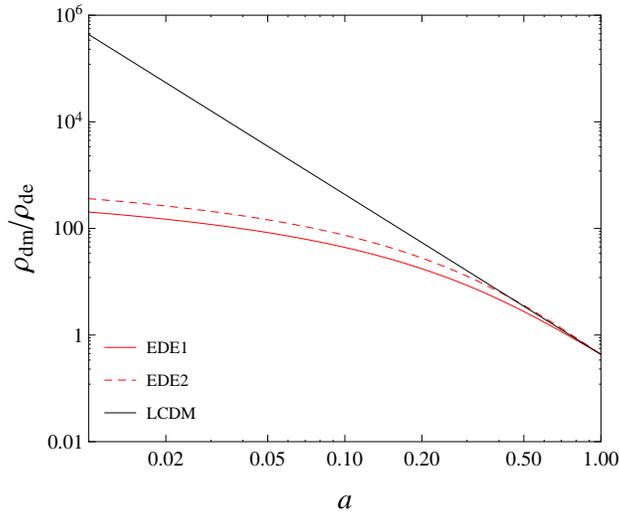}
\caption[width=1\columnwidth]{\label{dm2deEDE}The ratio of DM densities to DE densities
for different DE
models when there is no interaction between dark
sectors.  }
\end{figure}

In the spatially flat
Friedmann-Robertson-Walker(FRW) universe, the
evolutions of the energy densities of DE and DM
in the background spacetime are governed by
\begin{eqnarray}
\rho^{\prime}_{dm}+3\mathcal{H}\rho_{dm}=aQ_{dm}\nonumber\\
\rho^{\prime}_{de}+3\mathcal{H}(1+w)\rho_{de}=aQ_{de},
\end{eqnarray}
where $H$ is the Hubble constant and
$\mathcal{H}=aH$ with $a$ the scale factor of the
universe. $Q_\alpha$ indicates the interaction
between dark sectors, where the subscript
`$\alpha$' refers to `$dm$' or `$de$'
respectively. We show the evolution of the DE
fractional energy density when there is no interaction between
DE and DM in Figure \ref{bg_EDE_const}. In the left panel of Figure
\ref{bg_EDE_const}, we compare the $\Omega_{de}$ for
EDE1 with constant EoS DE and cosmological
constant. We can see that for the model
EDE1, DE started to have a significant ratio in the budget of the universe earlier,
which help to alleviate the coincidence
problem. We also compare the evolution of $\Omega_{de}$ for different
EDE models with that of the cosmological
constant in the right panel of Figure
\ref{bg_EDE_const}. The evolution of $\Omega_{de}$
shows that the model EDE1 is favorable to ease
the coincidence problem. To see more clearly, we present the behavior on the ratio $\rho_{dm}/\rho_{de}$ in Figure \ref{dm2deEDE}.
It is easy to see that the ratio for EDE1 is smaller at early times. This shows that the ratio for EDE1 evolves slower, so that it has longer period for the energy densities of EDE1 and DM to be comparable, in the spirit of alleviating the  coincidence problem.

Since we know the nature of neither DM
nor DE, it is hard to
describe the interaction between them, although
there are some attempts on this task
\cite{1202.0499, Sandro1, Sandro2, 1206.2589}.
Our study on the interaction between
dark sectors will concentrate on the
phenomenological descriptions. We assume there is energy
flow due to the interaction between dark sectors
where the coupling vector is defined in the form
$ Q^\nu=(\frac{Q}{a},0,0,0)^T $ \cite{1012.3904}
, and $Q$ takes the phenomenological form
$Q=3\lambda_{1}H\rho_{dm}$ or
$Q=3\lambda_{2}H\rho_{de}$, where  $\lambda_1$ and
$\lambda_2$ refer to the strength of the
respective couplings.

\begin{figure}[htp]
 \begin{tabular}{cc}
    \begin{minipage}[t]{2in}
    \includegraphics[width=1.0\columnwidth]{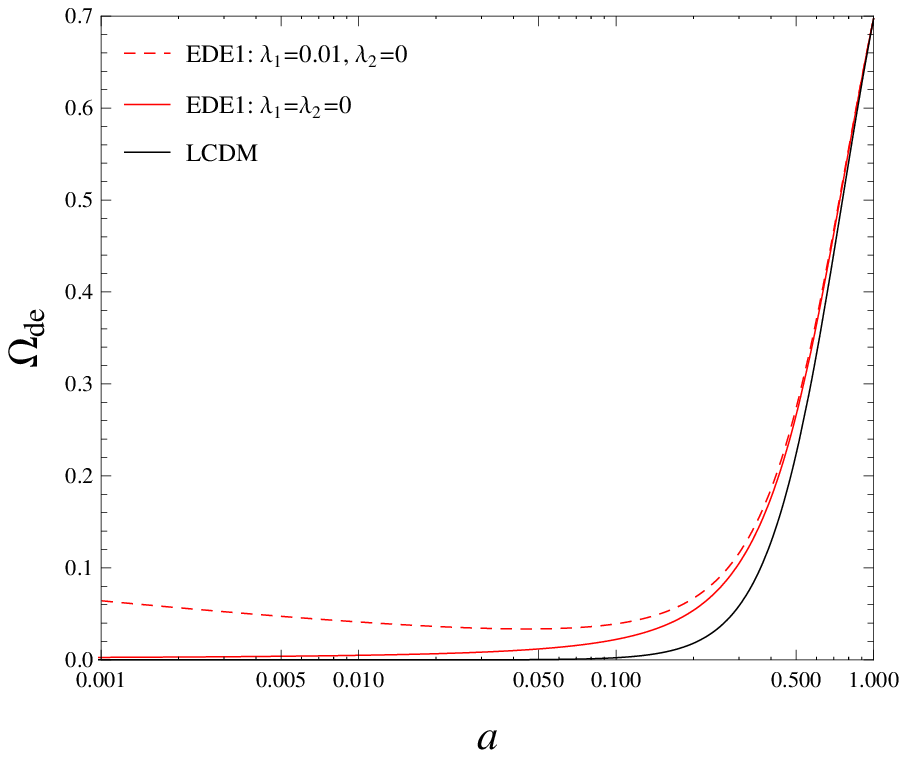}
    \end{minipage}
     \begin{minipage}[t]{2in}
    \centering
    \includegraphics[width=1\columnwidth]{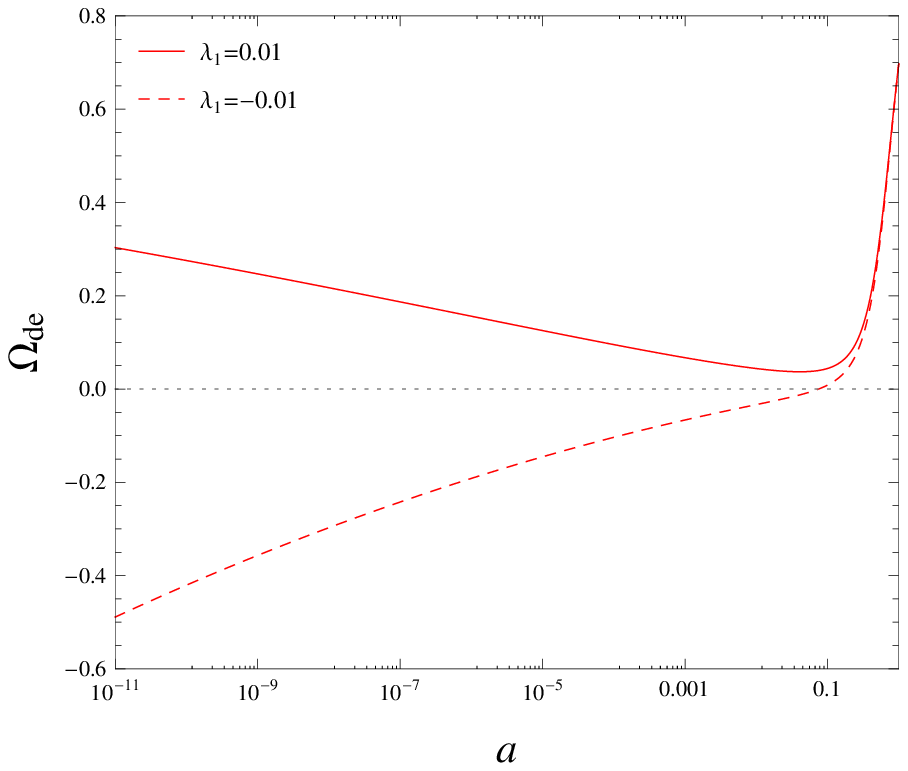}
      \end{minipage}
    \begin{minipage}[t]{2in}
    \centering
    \includegraphics[width=1\columnwidth]{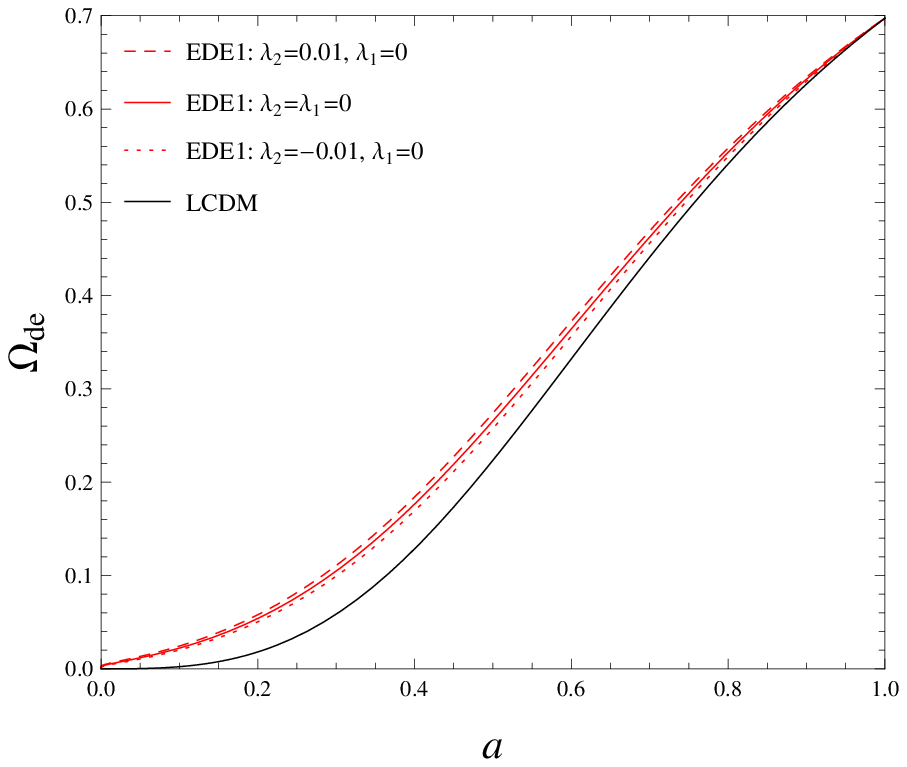}
    \end{minipage}
\end{tabular}
\caption{\label{bg_la}The evolutions of DE
fractional  densities when there is
interaction between dark sectors. }
\end{figure}

We plot the evolution of
the DE fractional energy density in Figure
\ref{bg_la}. In the left panel, we choose the
interaction as being proportional to the energy
density of DM.
For the EDE model, with the
positive coupling proportional to the DM energy
density ($\lambda_1 > 0$), the influence of DE in the universe
evolution appeared much earlier. The positive
coupling, in our notation, indicates that energy
flows from DE to DM
\cite{1311.7380,1308.1475,1012.3904,1001.0097,0710.1198}.
For the same amount of DE today, with the
positive coupling, it implies that DE density was
higher in the past.  The coupling strength $\lambda_1$ cannot be chosen negative, since the negative $\lambda_1$ will lead to the negative DE fractional energy density $\Omega_{de}$ at early time of the universe, as is shown in the middle panel of Figure \ref{bg_la}, which is certainly unphysical.  In the right panel of Figure \ref{bg_la}, we show
the case where the interaction is proportional to
the DE density. We see that for the EDE1 model
with positive coupling ($\lambda_2 > 0$), there was more
EDE at high redshift if the present DE amount is
the same as that of the $\Lambda$CDM model, what we
also argued to have consequences related to DM
phenomenology in accordance to results of BOSS
\cite{abdferreirawang}. But comparing with the
left panel, the influence of the coupling is
weaker in the right panel. This is easy to
understand, because in the right panel the
interaction is proportional to DE energy density,
which was much lower than that of DM at early
times in the universe.

Besides the background dynamics, we can
extend the study to the linear relativistic
evolution of the system of DE and DM. The gauge
invariant linear perturbation equations of the
system were derived
in\cite{0807.3471,0902.0660,0906.0677,1012.3904}.
Using the
phenomenological form of the energy transfer
between dark sectors defined above, the equations yield
\begin{eqnarray}
D_{dm}^{\prime}&=&-kU_{dm}+3\mathcal{H}\Psi{(\lambda_1+\lambda_2/r)}-3(\lambda_1+\lambda_2/r)\Phi^{\prime}+3\mathcal{H}\lambda_2(D_{de}-D_{dm})/r,\nonumber\\
U_{dm}^{\prime}&=&-\mathcal{H}U_{dm}+k\Psi-3\mathcal{H}(\lambda_1+\lambda_2/r){U_{dm}}; \nonumber\\
D_{de}^{\prime}&=&-3\mathcal{H}(C_e^2-w)D_{de}+\{3w^{\prime}-9\mathcal{H}(w-C_e^2)(\lambda_1{r}+\lambda_2+1+w)\}\Phi,\nonumber\\
&-&9\mathcal{H}^2(C_e^2-C_a^2)\frac{U_{de}}{k}+3(\lambda_1{r}+\lambda_2)\Phi^{\prime}-3\Psi\mathcal{H}(\lambda_1{r}+\lambda_2)+3\mathcal{H}\lambda_1{r}(D_{de}-D_{dm})\nonumber\\
&-&9\mathcal{H}^2(C_e^2-C_a^2)(\lambda_1{r}+\lambda_2)\frac{U_{de}}{(1+w)k}-kU_{de}, \nonumber\\
U_{de}^{\prime}&=&-\mathcal{H}(1-3w)U_{de}-3kC_e^2(\lambda_1{r}+\lambda_2+1+w)\Phi+3\mathcal{H}(C_e^2-C_a^2)(\lambda_1{r}+\lambda_2)\frac{U_{de}}{(1+w)}\nonumber\\
&+&3(C_e^2-C_a^2)\mathcal{H}U_{de}+kC_e^2D_{de}+(1+w)k\Psi+3\mathcal{H}(\lambda_1{r}+\lambda_2)U_{de}\quad ,\label{3}
\end{eqnarray}
where $\Psi, \Phi$ are gauge invariant gravitational potentials, $D_{\alpha} = \delta_{\alpha}-\frac{\rho'_{\alpha}}{\rho\mathcal{H}}\Phi$
is the gauge invariant density contrast, $U_{\alpha}= (1 + w_{\alpha})V_{\alpha}$, $V_{\alpha}$ is the gauge invariant
peculiar velocity, and $r\equiv \rho_{dm}/\rho_{de}$ is the
energy density ratio of DM and DE. $C_a$ is the
adiabatic sound speed of DE and $C_e$ is the effective
sound speed of DE which we will set to be $1$ in this
work. Having these perturbation equations,
we are in a position to discuss the evolutions of
DE and DM density perturbations.

\begin{figure}[htp]
 \begin{tabular}{cc}
    \begin{minipage}[t]{3in}
    \includegraphics[width=1\columnwidth]{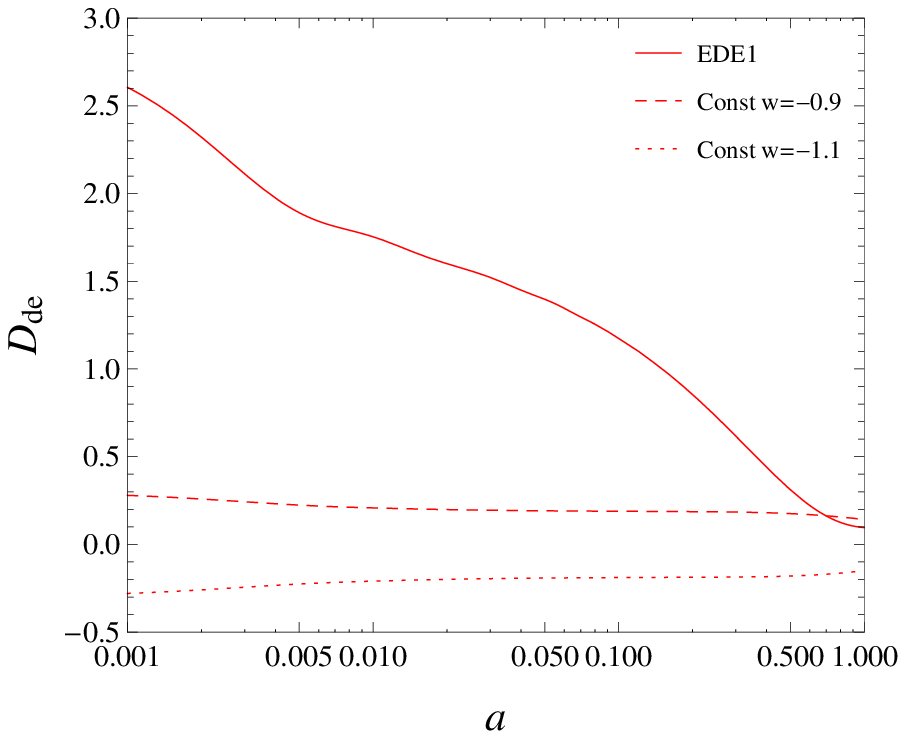}
    \end{minipage}
    \begin{minipage}[t]{3in}
    \centering
    \includegraphics[width=1.05\columnwidth]{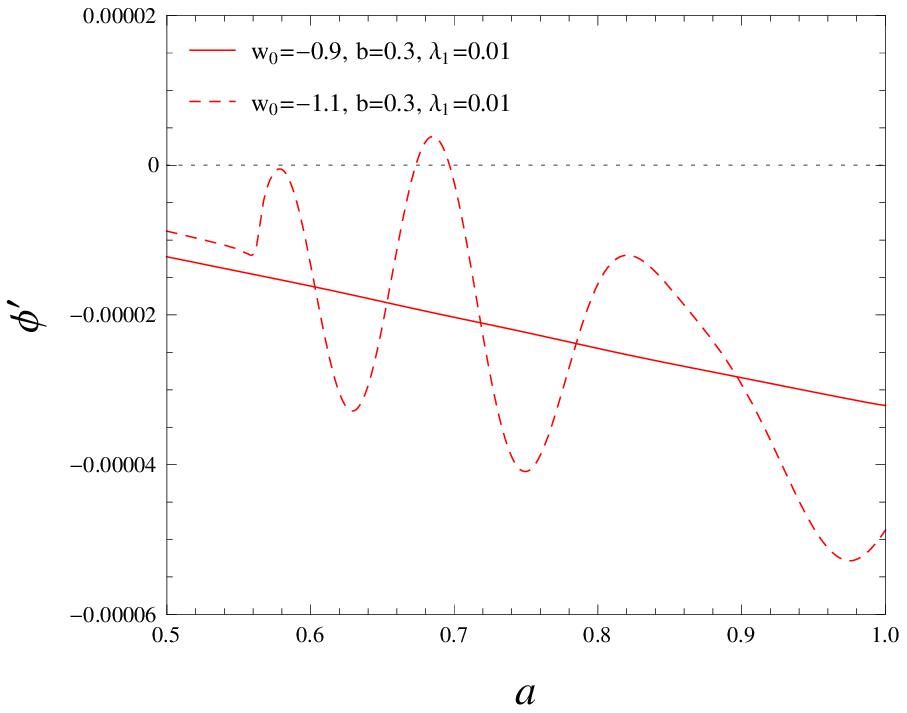}
    \end{minipage}
\end{tabular}
\caption{\label{DE_w}The left panel: The
evolutions of DE perturbations. The right panel: The time derivative of the gravitational potential. }
\end{figure}

\begin{figure}[htp]
\begin{center}
\subfigure[]{\label{DM_wa}\includegraphics[width=0.45\textwidth]{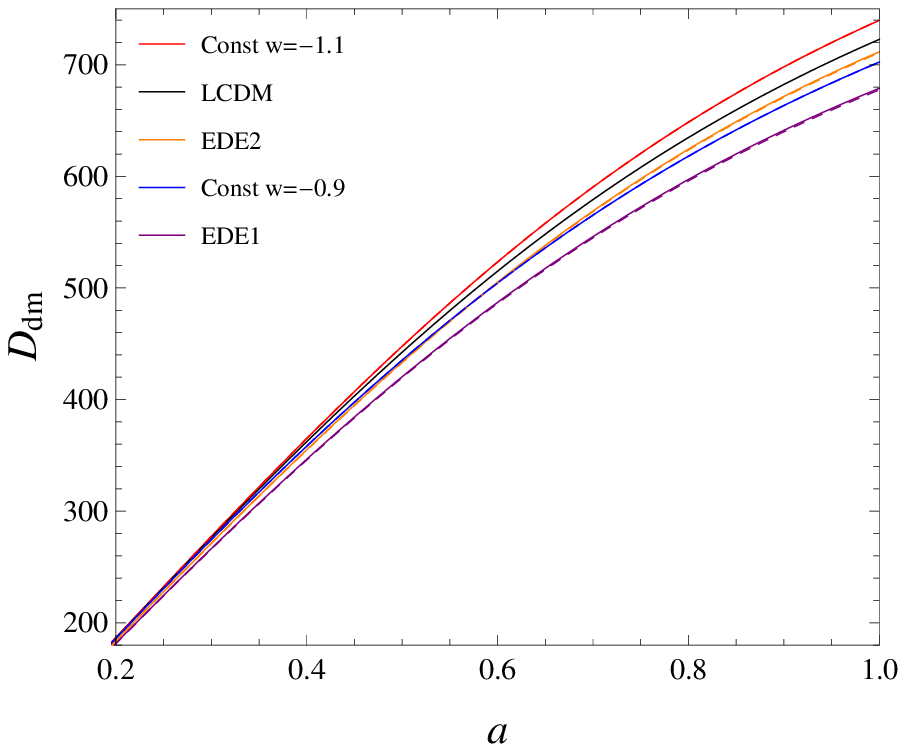}}
\subfigure[]{\label{DM_wb}\includegraphics[width=0.45\textwidth]{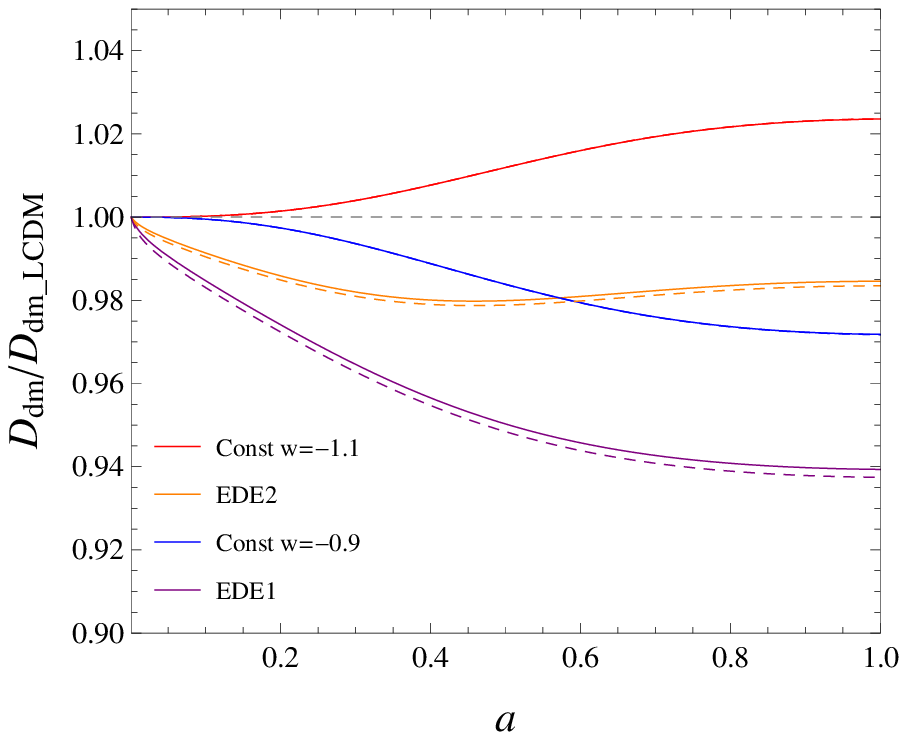}}
\subfigure[]{\label{DM_wc}\includegraphics[width=0.45\textwidth]{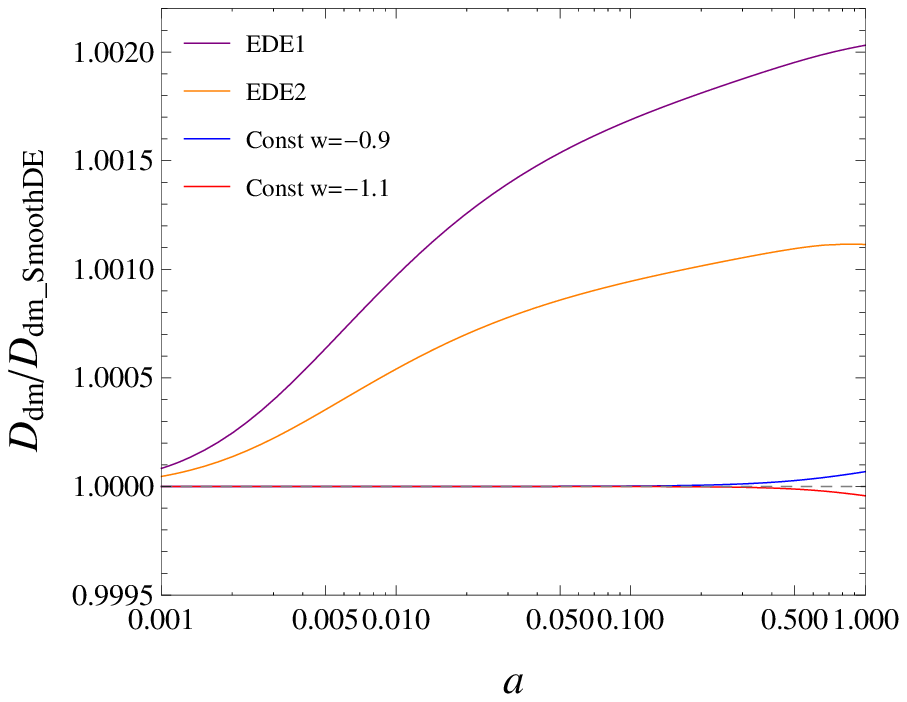}}
\caption{(a) The evolutions of DM perturbations.
(b) The comparison of DM perturbation evolutions
with that of $\Lambda$CDM model. (c) The
comparison of DM perturbation evolutions with and
without DE perturbations. The solid lines refer to models taking into account DE perturbations. The dashed lines refer to the models assuming homogenous DE. }
\end{center}
\end{figure}

Assuming $\lambda_1=\lambda_2=0$ in (\ref{3}), we
display the evolution of the DE perturbation in the left panel of
Figure \ref{DE_w}. In contrast to the DE models
with constant EoS, which always have very small
DE fluctuations, we see that although the
fluctuation of EDE decays to zero as its EoS
approaches to the cosmological constant, at early
times, when EDE started to play a significant
role, its fluctuation was not too small.  It
would be interesting to investigate how the EDE
perturbation influences the growth of DM perturbations. We
display the result in Figure \ref{DM_wa}, where
we show the evolution of DM density perturbation in different
DE models. It is clear that the earlier
presence of non-negligible DE fractional density in the
background suppresses the growth in the DM
perturbation. To see more closely, we have
compared the evolution of the DM perturbation to
the standard $\Lambda$CDM model in Figure
\ref{DM_wb}. DM perturbations were suppressed compared
with $\Lambda$CDM model if DE is
described by EDE2, constant
$w=-0.9$ and EDE1. The only exception is when DE has a constant EoS
$w=-1.1$. The difference
in the structure growth from that of the
$\Lambda$CDM model can be mainly attributed to
the differences in the background DE fractional
energy density from the standard $\Lambda$CDM
model. The suppression of the growth of perturbations was
caused by the excessive amount of DE than that in
the $\Lambda$CDM model at early epoch, which
hindered gravitational attraction  and
weakened the growth of DM perturbations. For the EDE models,
especially EDE1, the further excess of $\Omega_{de}$
at early times suppresses the structure growth even
more. The solid lines indicate the models having DE perturbation, while the dashed lines
are for the homogeneous DE models where the DE
perturbations are neglected. For the DE models
with constant EoS, the difference of effects on
the DM perturbations caused by homogeneous and
inhomogeneous DE are negligible. This can be
further seen in Figure \ref{DM_wc}. But for EDE models,
we clearly see the differences between the solid
and dashed lines for the inhomogeneous and
homogeneous DE.
Figure \ref{DM_wc} shows this property much clearer.
This is understandable because for the DE with
constant EoS,  the DE perturbation itself is
tiny. However for the EDE models, we clearly see
that different from the homogeneous DE model, the
DE perturbations do have impact on DM perturbations.

\begin{figure}[htp]
\begin{center}
\subfigure[]{\label{DM_INa}\includegraphics[width=0.45\textwidth]{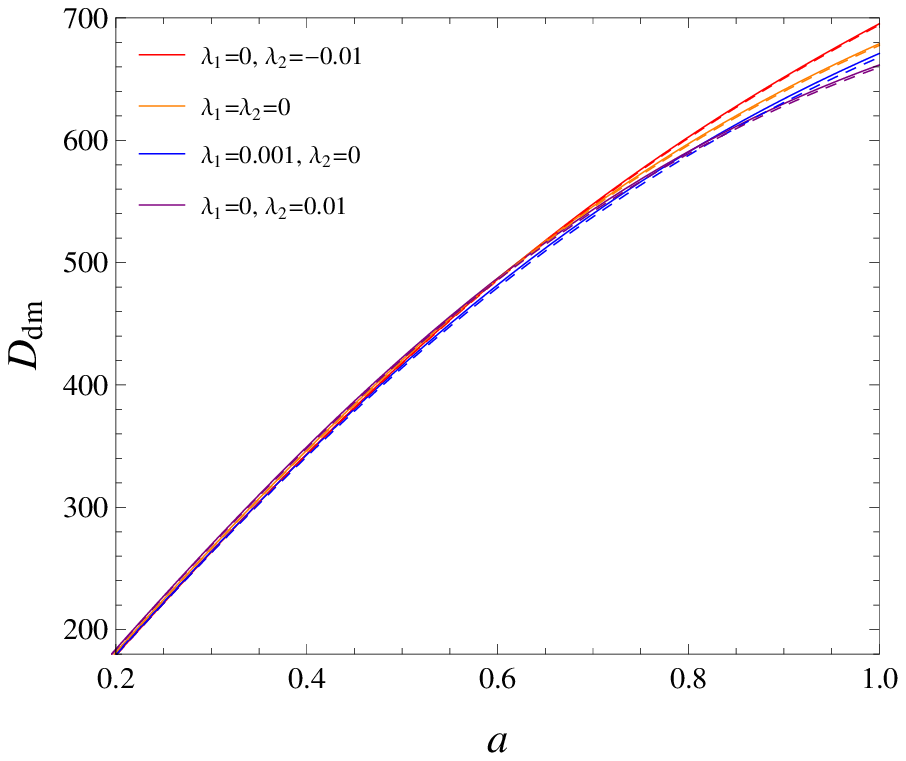}}
\subfigure[]{\label{DM_INb}\includegraphics[width=0.45\textwidth]{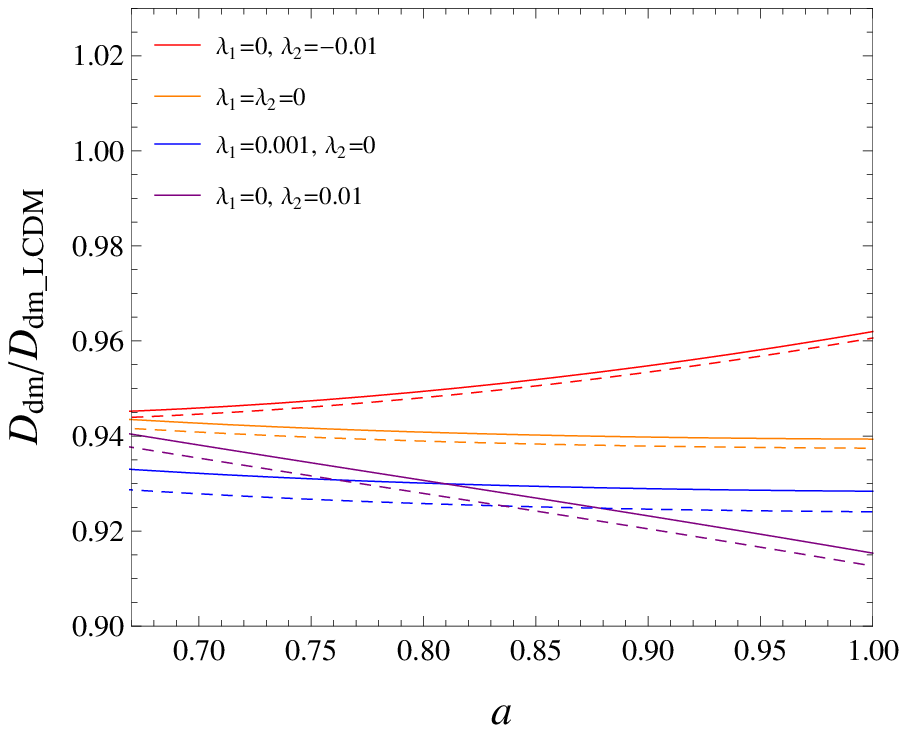}}
\subfigure[]{\label{DM_INc}\includegraphics[width=0.45\textwidth]{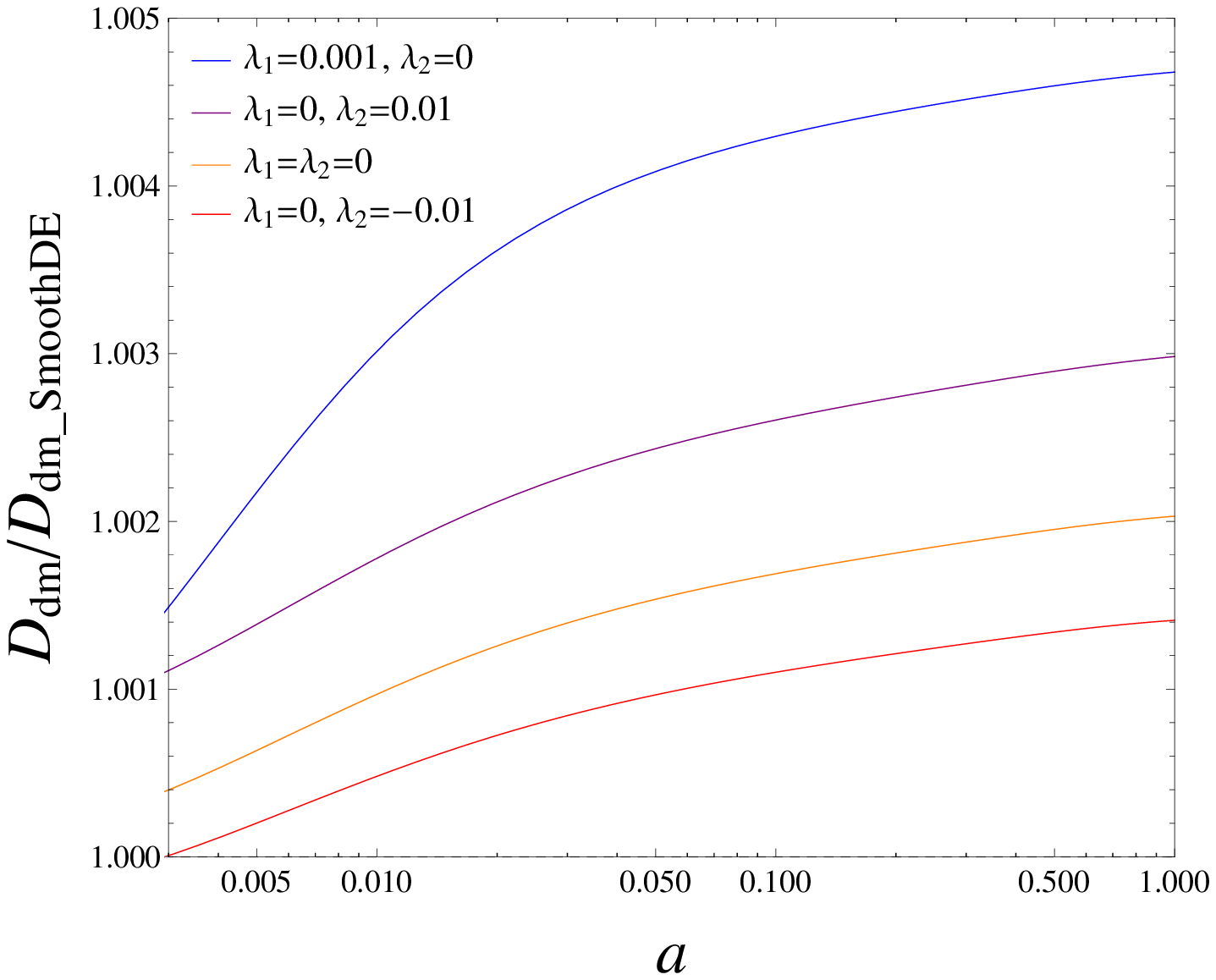}}
\caption{\label{DM_IN}(a) DM perturbations when
EDE interacts with DM. (b) The comparison of DM
perturbation evolutions with that of $\Lambda$CDM
model. (c) The comparison of DM perturbation
evolutions between models with and without DE perturbations. The solid lines refer to models taking into account DE perturbations. The dashed lines refer to the models assuming homogenous DE. }
\end{center}
\end{figure}

Considering the interaction between dark sectors,
the situation becomes more complicated. To see
clearly the influence of the interaction in dark
sectors on the linear perturbations, we
concentrate on the DE model EDE1,
because in EDE1 the DE EoS is always greater than $-1$. $1+w$ has to be positive in order to avoid oscillation in the time derivative of the gravitational potential on large scales if the interaction between dark sectors is proportional to the energy density of DM, which is inconsistent with observations, as shown in the right panel of Figure \ref{DE_w}.
In Figure
\ref{DM_INa}, we see that at the present moment for the
model with energy decay from EDE to DM, the DM
perturbation is smaller, which is different from
the case with energy decay in the opposite
direction. This is easy to understand, because
for the positive coupling the background
$\Omega_{de}$ was bigger in the past, which hindered
the structure growth. In Figure \ref{DM_INb}, we present the
comparison of the DM perturbations between the
interacting EDE model and the $\Lambda$CDM model.
It is clear that, comparing with $\Lambda$CDM
model, EDE interacting with DM leads to smaller DM
perturbations. For the interaction between dark sectors
proportional to the energy density of DM, the
effect of interaction showed up earlier. A positive $\lambda_1$ implies more
DE in the past, bringing further suppression in
the DM perturbations at early time. For the interaction
between dark sectors proportional to the density
of DE, the effect showed up later when DE
started to dominate. A positive $\lambda_2$
indicates the energy flow from EDE to DM, which implies
that there was more DE in the past,
preventing the DM perturbations further. This explains
why the line in Figure \ref{DM_INb} with positive
$\lambda_2$ is lower. For negative
$\lambda_2$, energy flows from DM to DE. To have the observed
amount of DM now, there must be more DM in the
past, which implies faster growth of DM perturbation.
Since this effect of interaction started to
appear when DE became important and became more
influential in the era of accelerated expansion, lines corresponding to positive and
negative $\lambda_2$ in Figure \ref{DM_INb} deviate from the non-interacting case late in the history of the universe.
The solid and dashed lines in Figure \ref{DM_INb}
refer to inhomogeneous and homogeneous DE,
respectively. We see that DM perturbations
differ by including the DE perturbations or not.
This can be seen much clearer in Figure \ref{DM_INc}.
Comparing with Fig.\ref{DM_wc}, we see that when
energy flows from DE to DM, the difference in the
DM perturbations caused by the inhomogeneous DE and
homogeneous DE is enlarged. Also from Figure \ref{DM_INc}
we see that the difference in the DM perturbations
between inhomogeneous and homogeneous
DE is more sensitive to the coupling if it is
proportional to the energy density of DM.

\section{CMB power spectrum}

Once we have the understanding of the linear
perturbations for DM and DE, we can proceed to
study the effects of DE models on CMB. On
large scales, the CMB power spectrum is composed
of the ordinary Sachs-Wolfe (SW) effect and the
Integrated SW (ISW) effect. The SW effect
indicates the photons' initial condition when it
left the last scattering surface while the ISW
effect is the contribution due to the change of
the gravitational potential when photons passing
through the universe on their way to earth. The
gauge invariant gravitational potential in the
absence of anisotropic stress can be given by the
Poisson equation $k^2\Phi=-4\pi G a^2\delta\rho$.
Its derivative in DM plus DE universe,
which is the source term for the ISW contribution,
is given by $k^2\Phi'=-4\pi
G\frac{\partial}{\partial \eta
}[a^2(\delta\rho_{dm}+\delta\rho_{de})]$. Thus the
large scale CMB power spectrum depends on the evolution
of the density perturbations of DE and DM.
However it should be noted that ISW effect is
complicated. Besides density perturbations in DM
and DE, other cosmological parameters such as the
EoS of DE, background energy densities and $H_0$
etc. also have influence on it. Only for the same
background evolution, the large scale CMB power
spectrum can be interpreted in terms of the
evolution of the density perturbations for DE and
DM.

\begin{figure}[htp]
\begin{center}
\subfigure[]{\label{cl_consta}\includegraphics[width=0.45\textwidth]{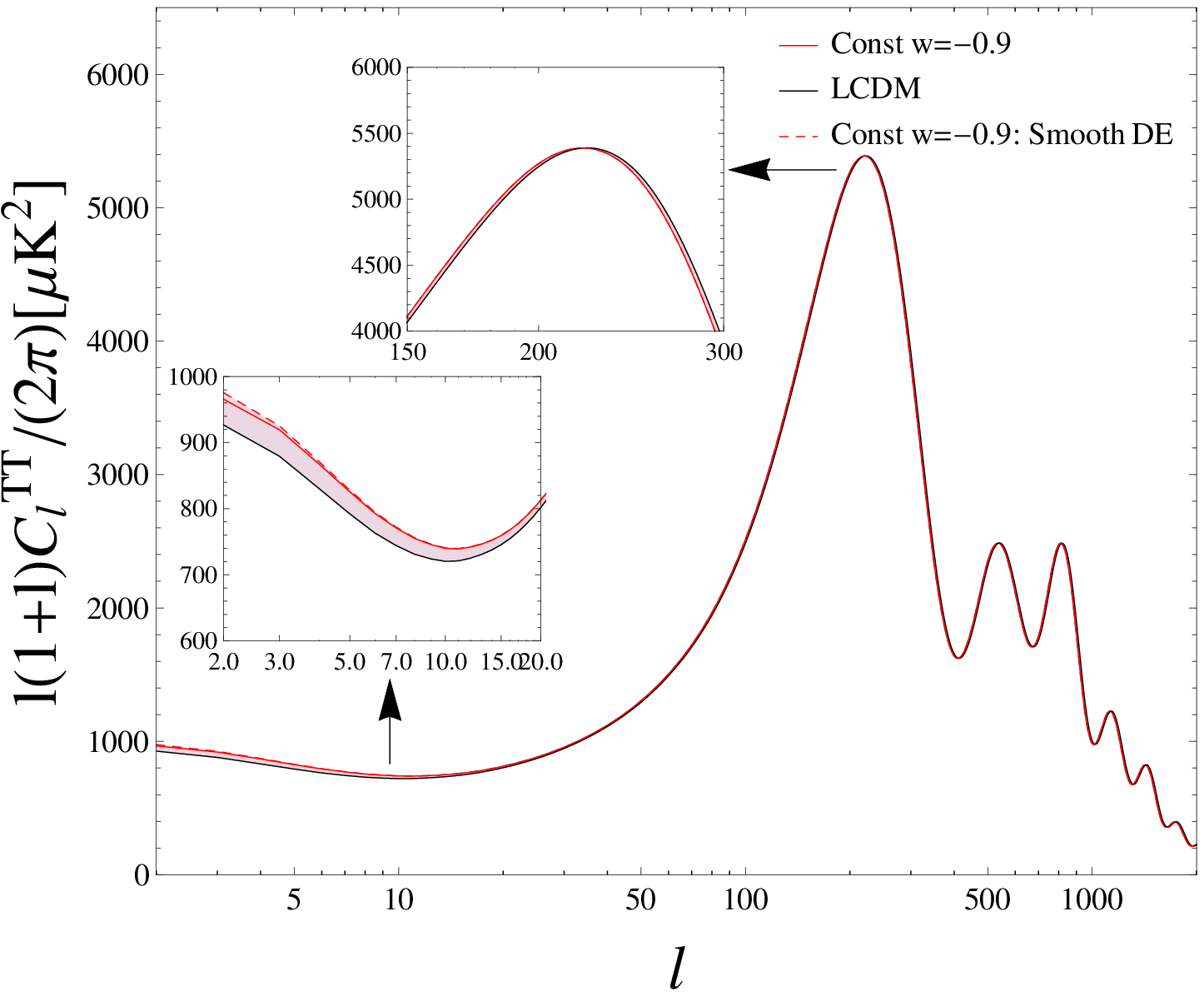}}
\subfigure[]{\label{cl_constb}\includegraphics[width=0.465\textwidth]{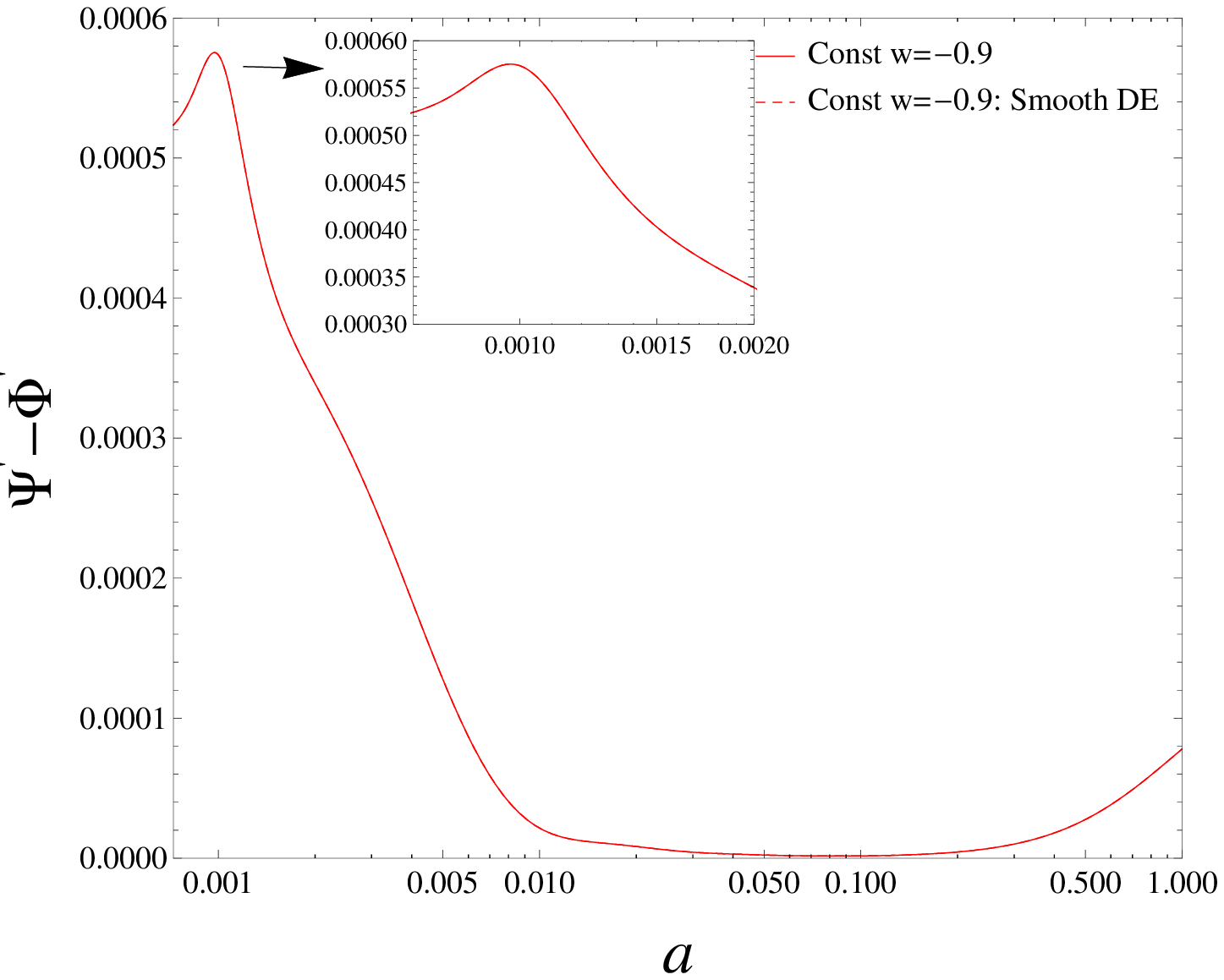}}
\caption{\label{cl_const}CMB power spectrum and the time derivative of the gravitational
potential when DE EoS is constant, $w=-0.9$}
\end{center}
\end{figure}

Neglecting the interaction between dark sectors,
for DE with constant EoS $w=-0.9$,
we show the CMB power spectrum in Figure
\ref{cl_consta}. Comparing with the $\Lambda$CDM
model, there is little difference in CMB at small
$l$ ISW effect. The ISW effect relates
to the time variation of the gravitational potentials,
which demonstrates little difference
between DE with constant EoS and $\Lambda$CDM \cite{xx}. For DE with constant EoS,
the CMB power spectrum keeps the same no
matter we include the DE fluctuations in the
computation or not.  The DE fluctuations do not
show up in the CMB power spectrum.  This is because for
DE with constant EoS, the DE
perturbation is negligible. And the result at
large scale CMB power spectrum agrees with what
disclosed in the  growth of DM perturbation in
the previous section where the DE fluctuations do
not show up for the DE with constant EoS. Thus
including the DE fluctuations, the CMB power spectrum
remains the same as when the
perturbations to DE is not taken into account.

\begin{figure}[htp]
\begin{center}
\subfigure[]{\label{cl_EDE1a}\includegraphics[width=0.45\textwidth]{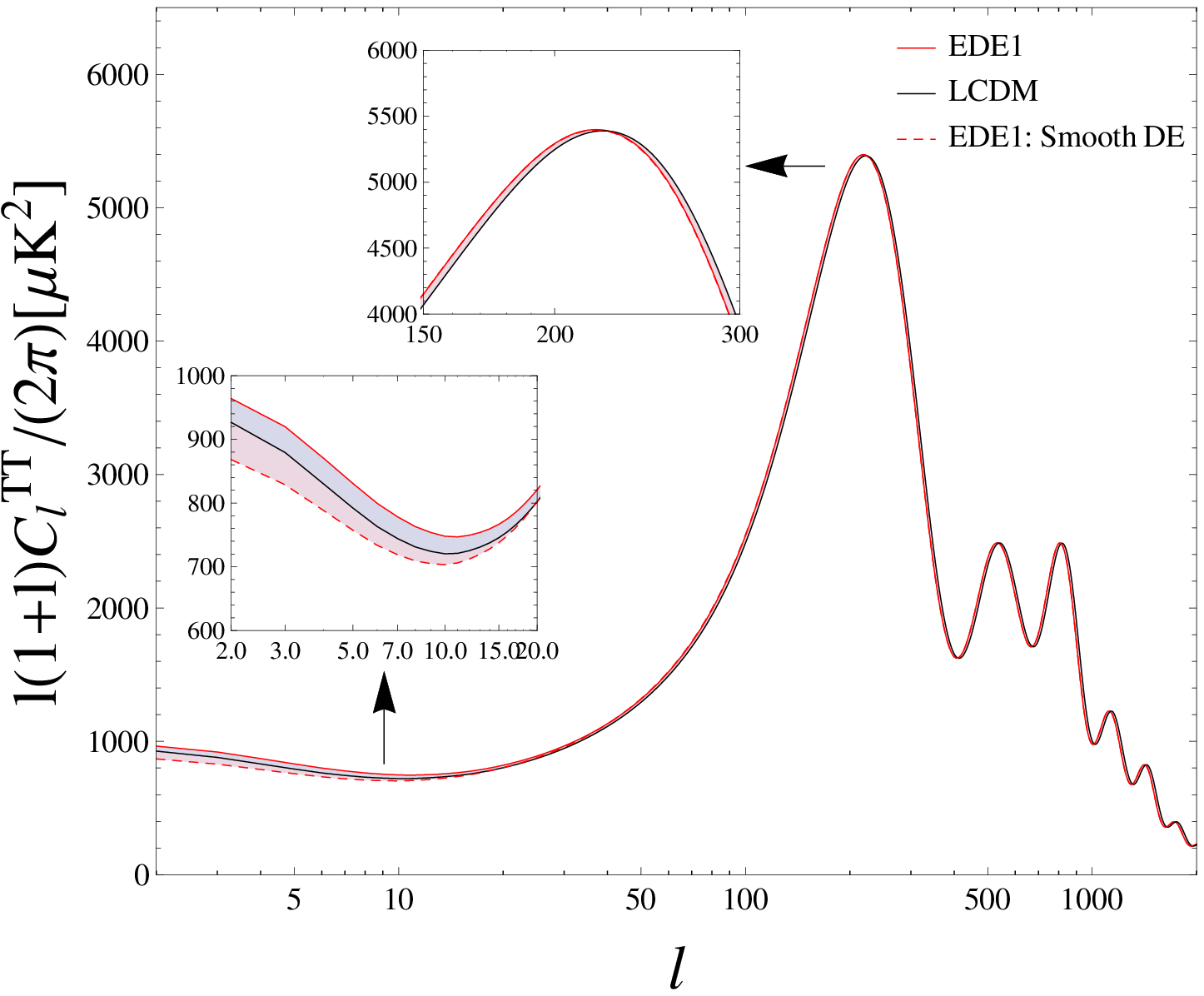}}
\subfigure[]{\label{cl_EDE2a}\includegraphics[width=0.45\textwidth]{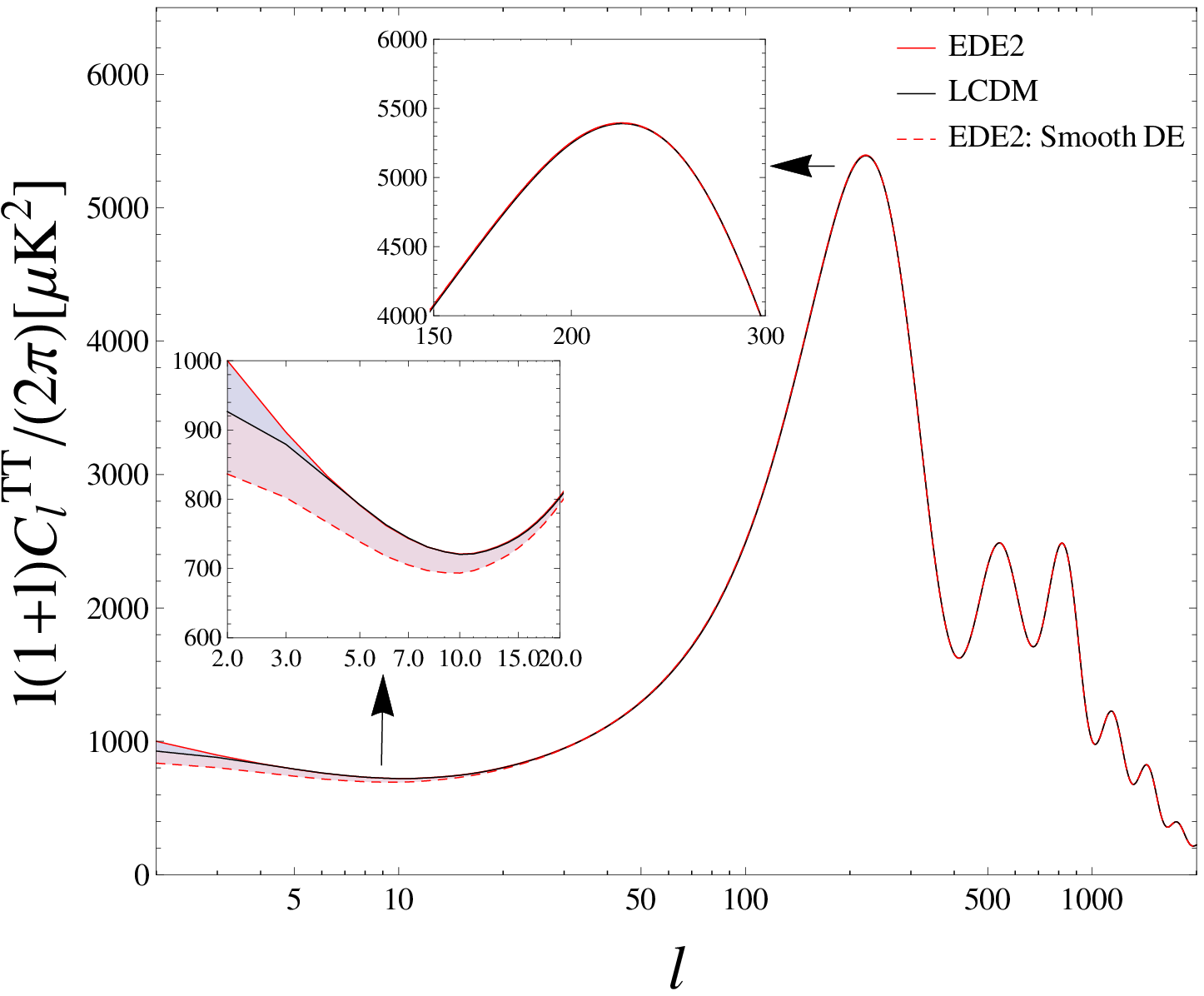}}
\caption{\label{cl_EDE1}CMB power spectrum for EDE
models. We compare
the CMB power spectrum for the universe with
inhomogeneous DE and homogeneous DE. For EDE models, the solid lines refer to inhomogeneous DE and the dashed lines refer to homogeneous DE.}
\end{center}
\end{figure}

For the EDE models, we observed some interesting
results in CMB. We considered both cases where EDE is homogeneous
and inhomogeneous in Figures \ref{cl_EDE1}. Besides a slight shift of the position of the acoustic peaks with respect to $\Lambda$CDM model, we see that the CMB power spectrum at small
$l$ is different between inhomogeneous and homogeneous EDE. In
homogeneous EDE model in which DE fluctuations is neglected,
the small $l$ spectrum is suppressed as compared
with that of $\Lambda$CDM. In inhomogeneous EDE model in which DE fluctuations is taken into account,
we observe an enhanced power spectrum at low $l$
with respect to $\Lambda$CDM. For a given EDE
model, the evolutions of background cosmological
parameters are the same, the differences in the
large scale CMB power spectrum can be attributed to the
evolutions of DE and DM density perturbations. In
the last section, we learnt that the
inhomogeneity in EDE will have an impact on the
DM perturbations. For the inhomogeneous EDE, the DM
perturbation is stronger. The inhomogeneous EDE
perturbation also evolves with time. These
effects result in a change of the gravitational
potential and the varition of the gravitational
potential in time  leads to the differences in
the ISW effect in CMB.

\begin{figure}[htp]
\begin{center}
\subfigure[]{\label{cl_la1a}\includegraphics[width=0.45\textwidth]{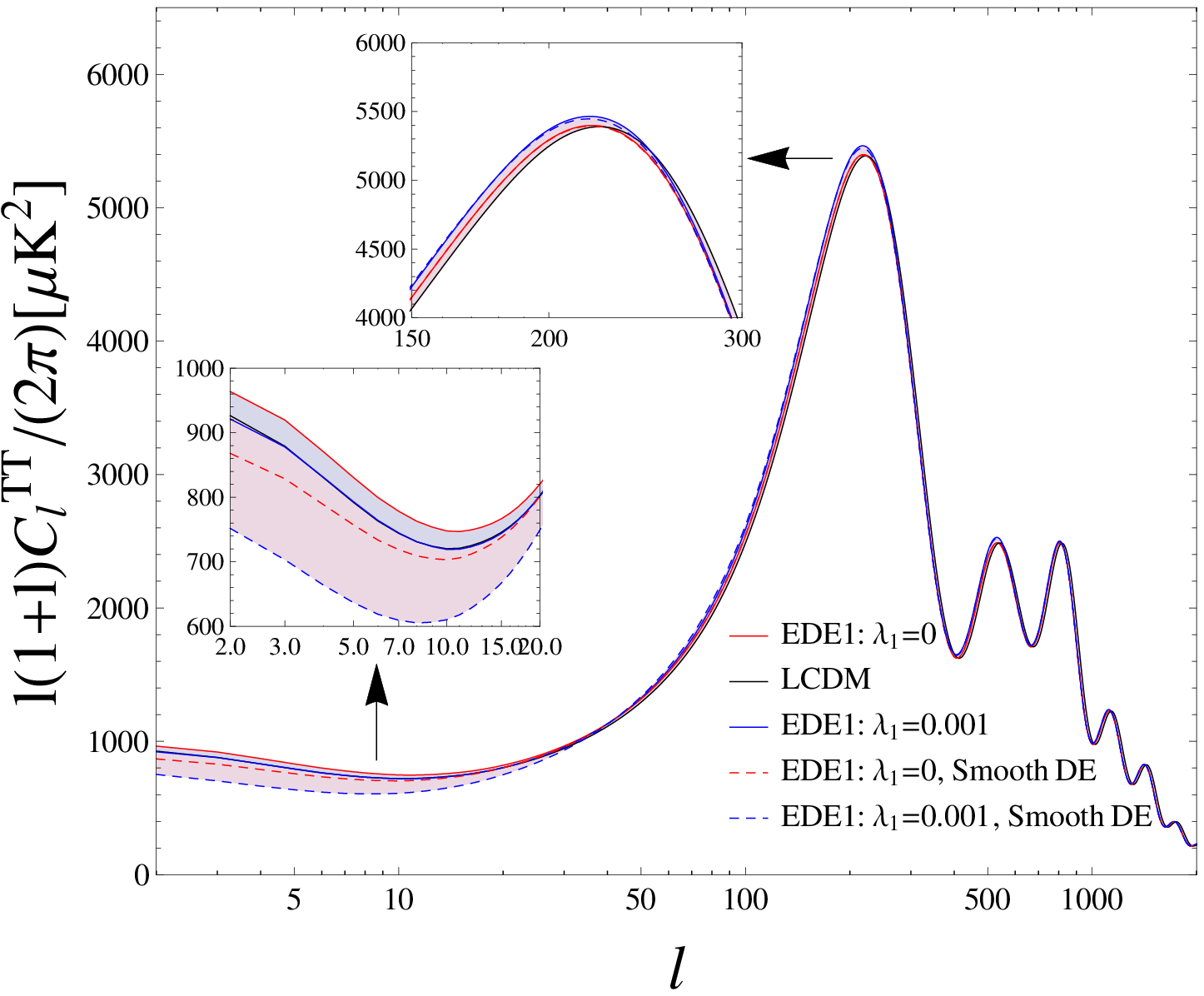}}
\subfigure[]{\label{cl_la2a}\includegraphics[width=0.45\textwidth]{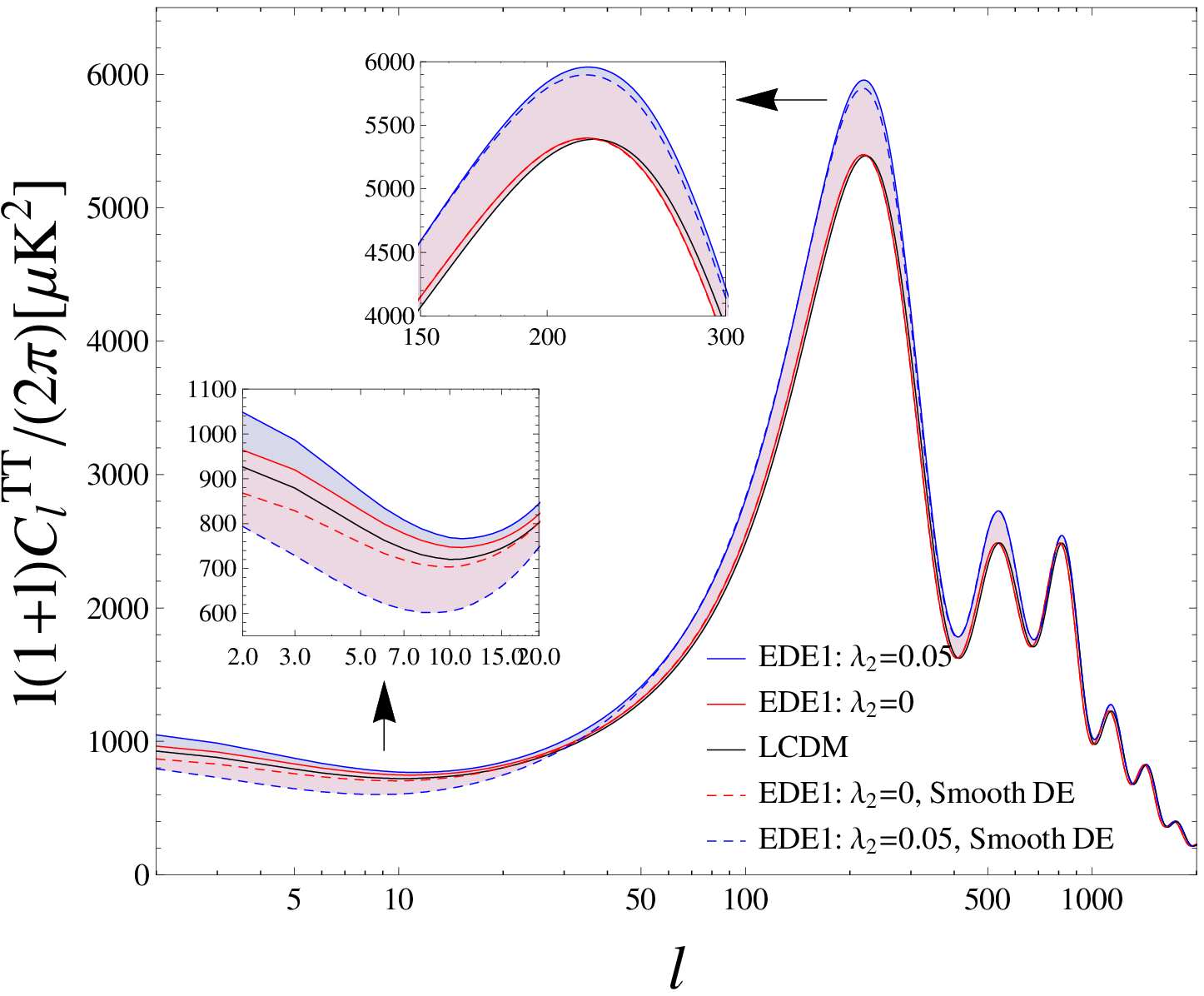}}
\caption{\label{cl_la1}CMB power spectrum for EDE
coupled to DM. For EDE models, the solid lines refer to inhomogeneous DE and the dashed lines refer to homogeneous DE. }
\end{center}
\end{figure}

Including interaction between dark sectors, we
have a richer physics in CMB. In the left panel
of Figure \ref{cl_la1}, we present the CMB power
spectrum for the interaction proportional to the
energy density of DM. With a positive
interaction, we see that the difference at low
$l$ CMB between homogeneous and inhomogeneous
EDE is enlarged compared with the zero coupling
case.  This can be attributed to the enlarged
differences in the EDE perturbations together
with the DM perturbations between including
the EDE fluctuations or not in the presence of
the interaction between dark sectors. Besides the
difference we observe at low $l$, at the
first peak the differences between homogeneous
and inhomogeneous EDE for the
same coupling is small. In the right panel we
show the influence of interaction proportional to
the energy density of DE. With inhomogeneous EDE,
the interaction makes the power spectrum higher
(the blue solid line) at low $l$. But with
homogeneous EDE, the power spectrum at small $l$
is suppressed (the blue dashed line). The
interaction between dark sectors enlarge the
differences in the small $l$ CMB power spectrum between homogeneous and inhomogeneous EDE. Making
the strength of the interaction stronger
($\lambda_2=0.05$), we see the clear enhancement
of the first peak.

\begin{figure}[htp]
\begin{center}
\subfigure[]{\label{cl_lasa}\includegraphics[width=0.45\textwidth]{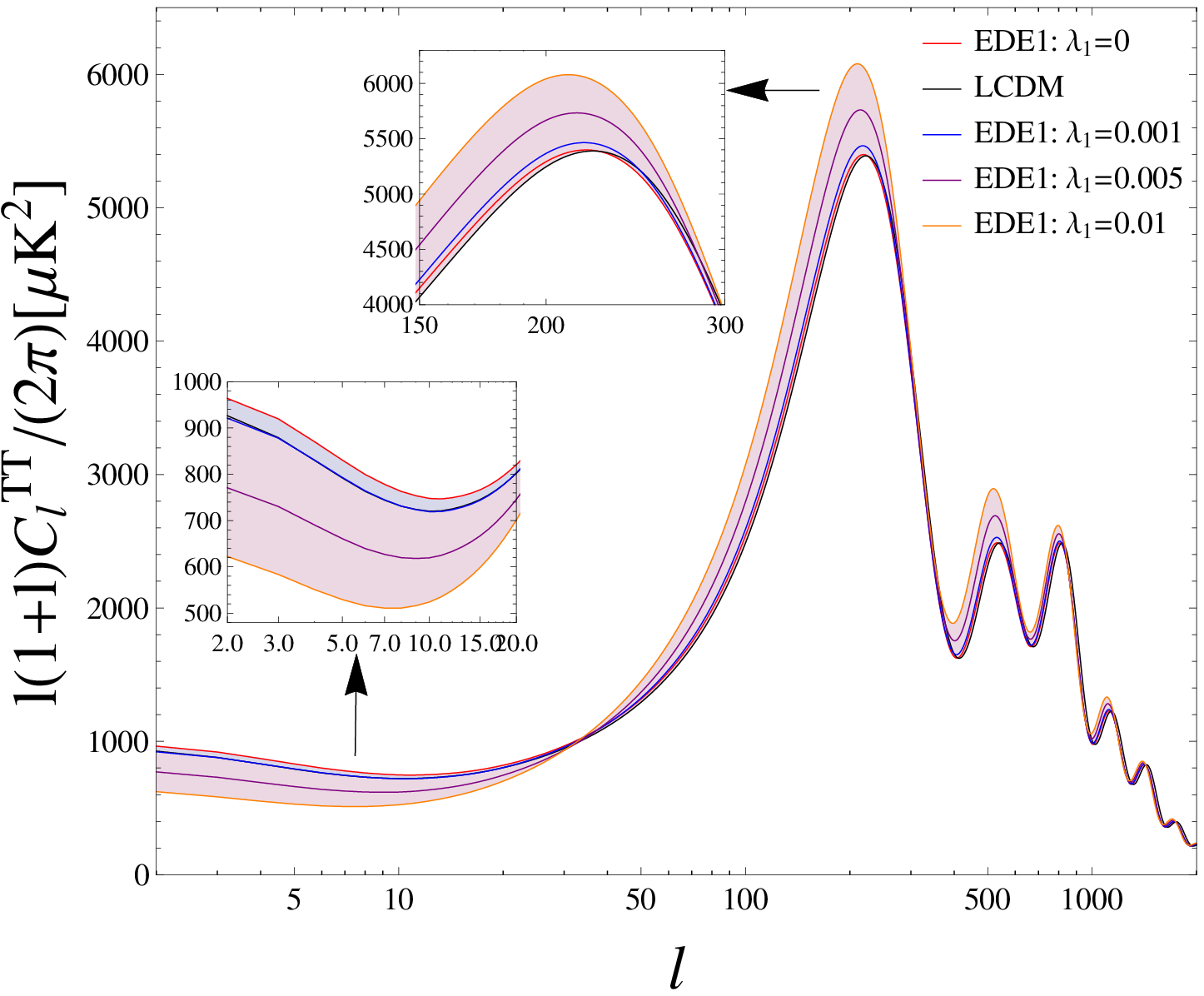}}
\subfigure[]{\label{cl_lasb}\includegraphics[width=0.45\textwidth]{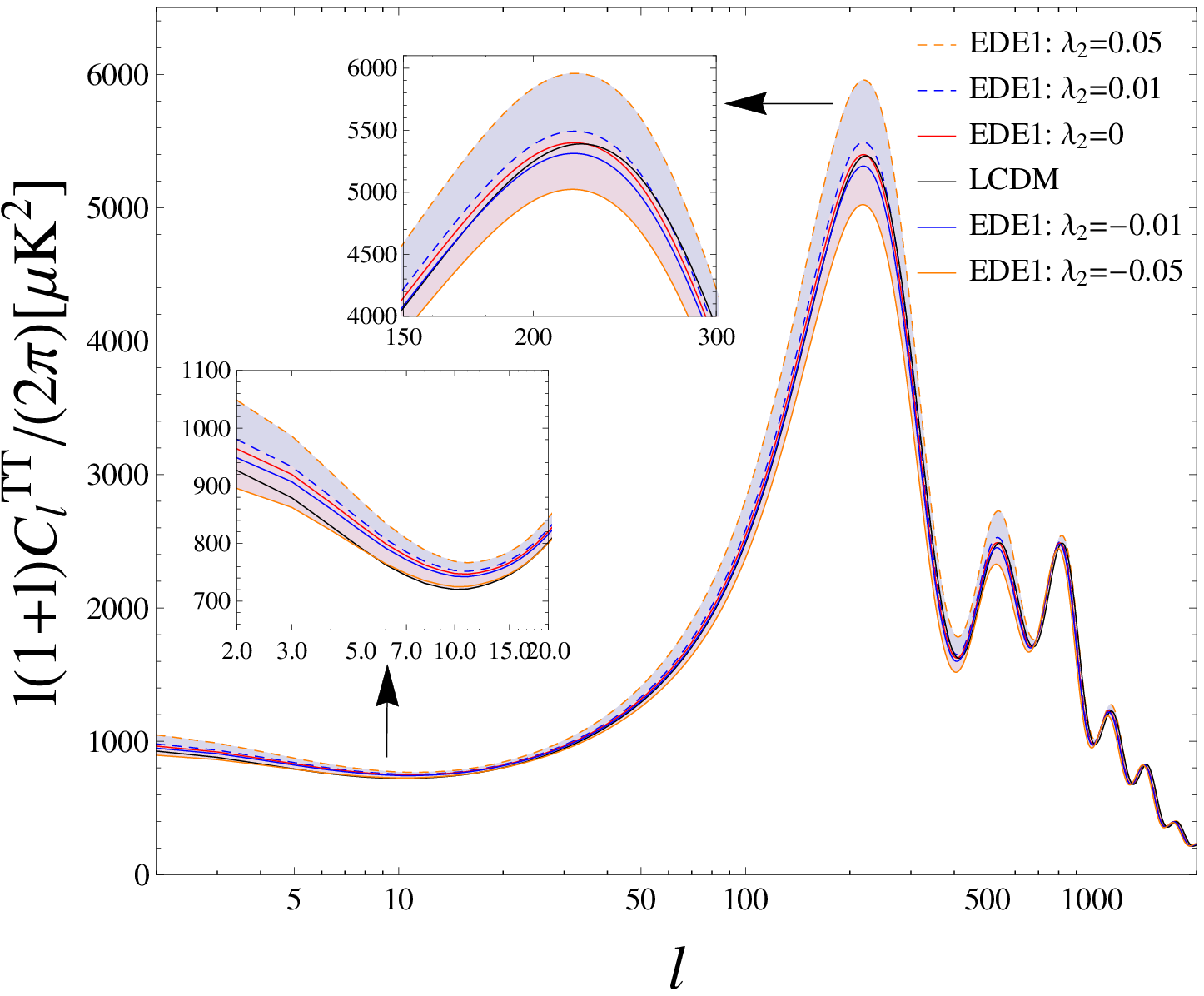}}
\caption{\label{cl_las}CMB power spectrum for EDE1 coupled to DM.
For EDE models, the solid lines refer to inhomogeneous DE and the dashed lines refer to homogeneous DE. }
\end{center}
\end{figure}

To disclose the influence of different forms and strength of the
interaction , we show the CMB power spectrum for EDE1 with various interactions as well as $\Lambda$CDM in Figure
\ref{cl_las}. We see that when the interaction is
proportional to the density of DM,  its influence appears not only at small $l$ but also at the
first acoustic peak of the CMB power spectrum. A larger $\lambda_1$
accommodates the suppression at low $l$ spectrum
but also the enhancement of the first peak. If
the interaction is proportional to the energy
density of DE, the CMB power spectrum exhibits
consistent behaviors both at low $l$ and the first
peak: a larger $\lambda_2$ leads to the
enhancement of the power spectrum. For the EDE models, the influences
of the interaction between dark sectors present
the same qualitative influence on CMB as compared
with the DE with constant EoS
\cite{0906.0677, 1012.3904}.

\section{Fitting results}
In this section we fit the EDE models to observations by Markov chain Monte Carlo (MCMC) method. We modify the public code CosmoMC\cite{cosmomc1}\cite{cosmomc2} to perform the MCMC analysis.
For the EDE models without interaction with DM, we carry out the fittings using two datasets: the CMB observations from
$Planck$(TT+TE+BB+EE)\cite{planck1,planck2,planck3} and a combined dataset of
$Planck$(TT+TE+BB+EE)+BAO\cite{bao1}\cite{bao2}\cite{bao3}+SN\cite{SN}+$H_0$\cite{H0}. We try to use these
observational data to distinguish between
homogeneous and inhomogeneous EDE models. When there is interaction between EDE and DM, we fit the EDE models to the combined dataset only, since the CMB data alone cannot constrain the cosmological parameters tightly due to the degeneracy between the coupling strength and DE EoS. In our numerical fittings, the priors of the
cosmological parameters are listed in Table \ref{prior}. For the interaction proportional to the energy density of DM, we have to put strict limits on the priors of the parameters. In order to avoid the negative DE energy density in the early background dynamics and the oscillatory behavior in the time derivative of the gravitational potential, it is necessary that $\lambda_1>0$ and $w_0>-1$, as we discussed above. For the sake of consistency, we set the prior of $w_0$ bigger than $-1$ in all models. Besides, $b$ cannot be close to zero if $\lambda_1>0$. Otherwise the DE EoS would be approximately a constant above and close to $-1$ in the early time of the universe, as shown in Figure \ref{w_diff_EDE}, which is known to cause unstable growth of curvature perturbation\cite{0807.3471,Xu2011}. Furthermore, considering that CMB power spectrum is more sensitive to $\lambda_1$ than $\lambda_2$, following \cite{1311.7380}, the prior of $\lambda_1$ is tighter than $\lambda_2$.

\begin{table}[htp]
\centering
\caption[width=1\columnwidth]{\label{prior}The priors for cosmological parameters. $b$(I) refers to the prior of the parameter $b$ of EDE model for no interaction between DM and DE and the interaction proportional to the energy density of DE. $b$(II) refers to the prior for the interaction in proportional to the energy density of DM.}
\renewcommand\arraystretch{1.0}
\begin{tabular}{|p{3cm}<{\centering}|p{10cm}<{\centering}|}
\hline
\hline
Parameter & Prior \\
\hline
$\Omega_bh^2$ & [0.005, 0.1 ]\\
\hline
$\Omega_ch^2$ & [0.001, 0.5 ]\\
\hline
$100\theta$ & [0.5, 10]\\
\hline
$\tau$ & [0.01, 0.8]\\
\hline
$n_s$ & [0.9, 1.1]\\
\hline
$log(10^{10}A_s)$ & [2.7, 4]\\
\hline
$w_0$  & [-0.99, -0.3]\\
\hline
$b(I)$    & [0.0, 1] \\
\hline
$b(II)$    & [0.1, 1] \\
\hline
$\lambda_1$ & [0.0, 0.01]\\
\hline
$\lambda_2$ & [-0.5, 0.5]\\
\hline
\end{tabular}

\end{table}

\begin{figure}[htp]
      \begin{tabular}{cc}
    \begin{minipage}[t]{3in}
    \centering
    \includegraphics[width=1\columnwidth]{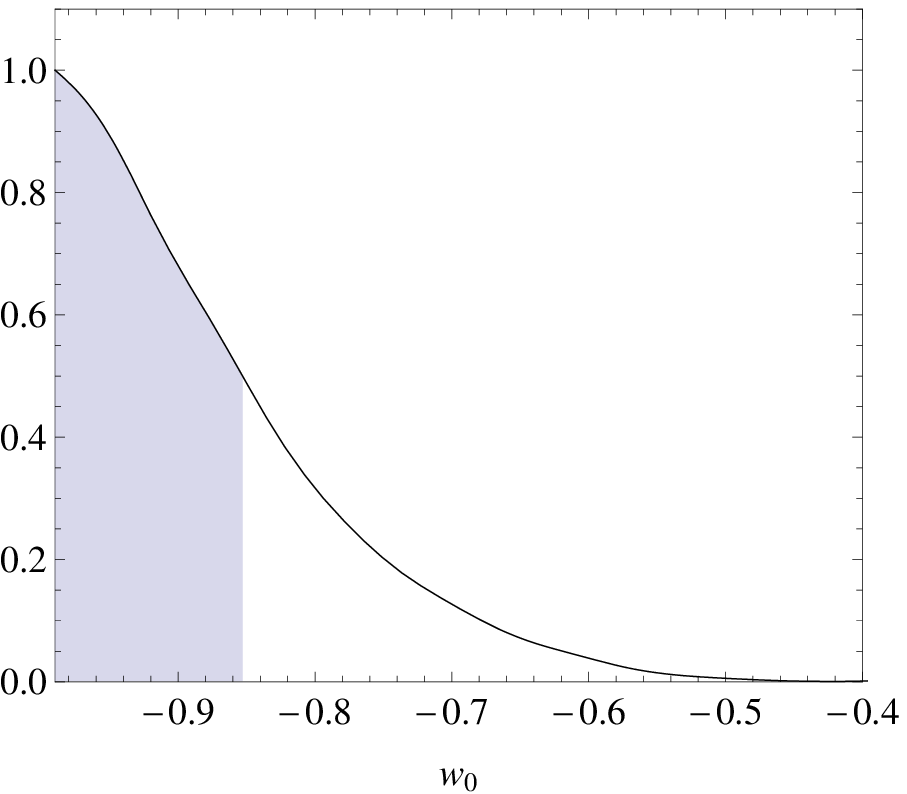}

    \end{minipage}
    \begin{minipage}[t]{3in}
    \centering
    \includegraphics[width=1\columnwidth]{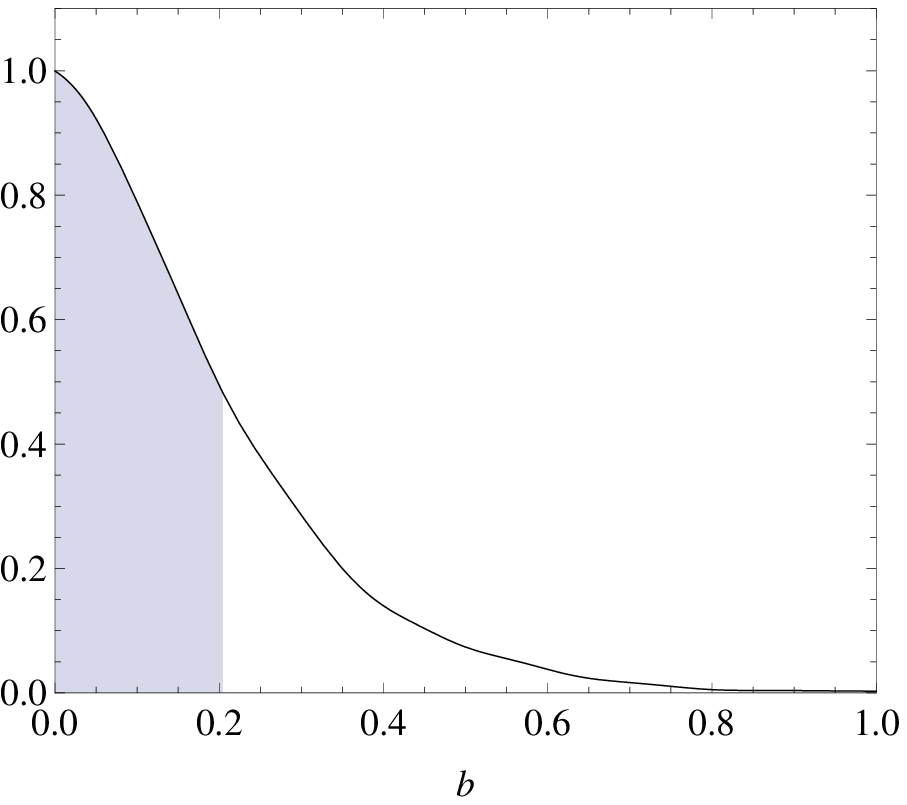}

    \end{minipage}
\end{tabular}
      \begin{tabular}{cc}
    \begin{minipage}[t]{3in}
    \centering
    \includegraphics[width=1\columnwidth]{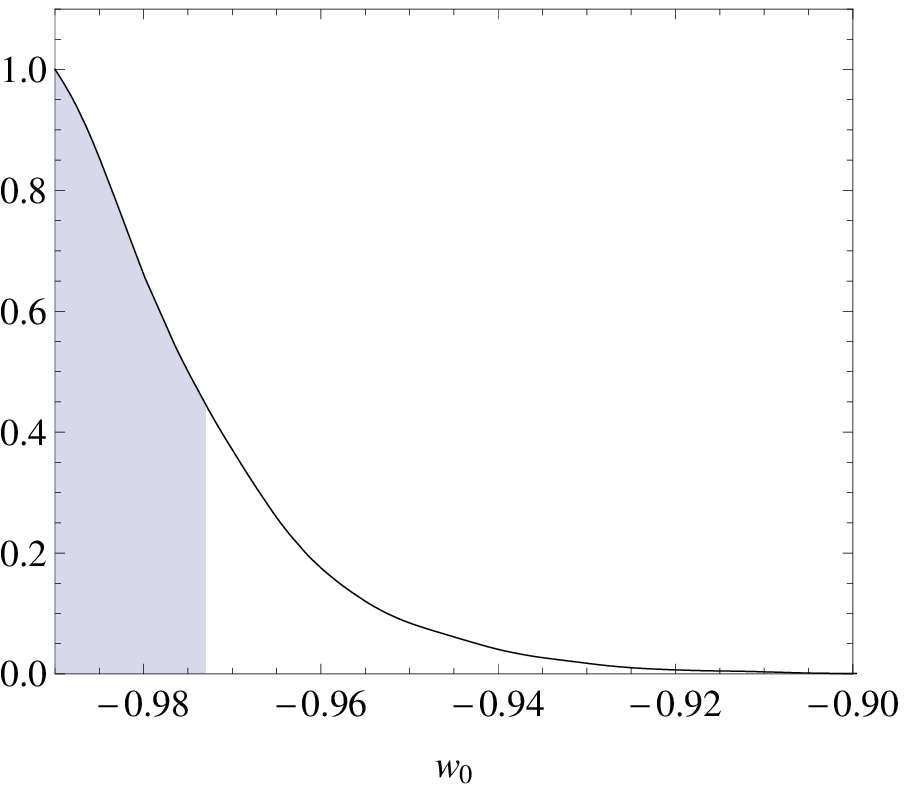}

    \end{minipage}
    \begin{minipage}[t]{3in}
    \centering
    \includegraphics[width=1\columnwidth]{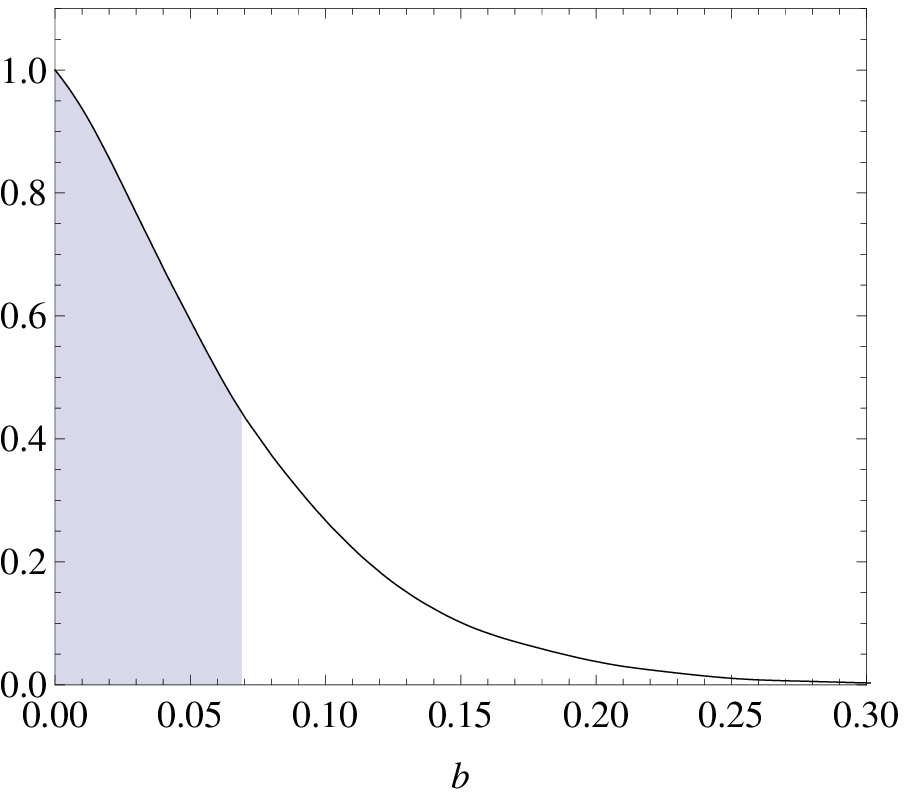}

    \end{minipage}
\end{tabular}
\caption{\label{Model_1}Fitting results of
the inhomogeneous EDE model. The upper panel is
from the {\it Planck} data alone, while the lower panel
is from the combined dataset.}
\end{figure}

 \begin{figure}[htp]
 \begin{tabular}{cc}
    \begin{minipage}[t]{3in}
    \includegraphics[width=1\columnwidth]{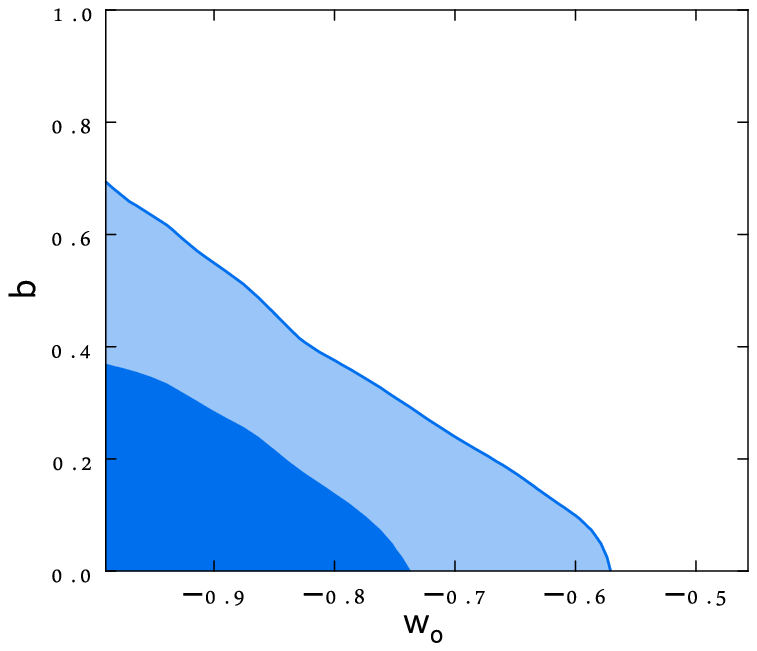}
    \end{minipage}
    \begin{minipage}[t]{3in}
    \centering
    \includegraphics[width=1\columnwidth]{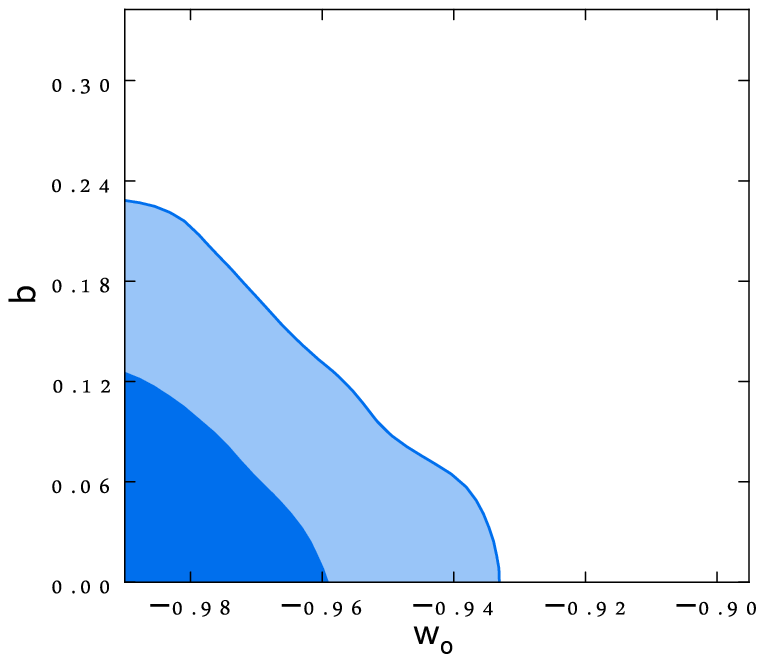}

    \end{minipage}
\end{tabular}
\caption{\label{2d}Fitting results of
the inhomogeneous EDE model in 2D contour $w_0 - b$. The left panel is
from the {\it Planck} data alone, while the right panel
is from the combined dataset.  }
\end{figure}

\begin{figure}[htp]
      \begin{tabular}{cc}
    \begin{minipage}[t]{3in}
    \centering
    \includegraphics[width=1\columnwidth]{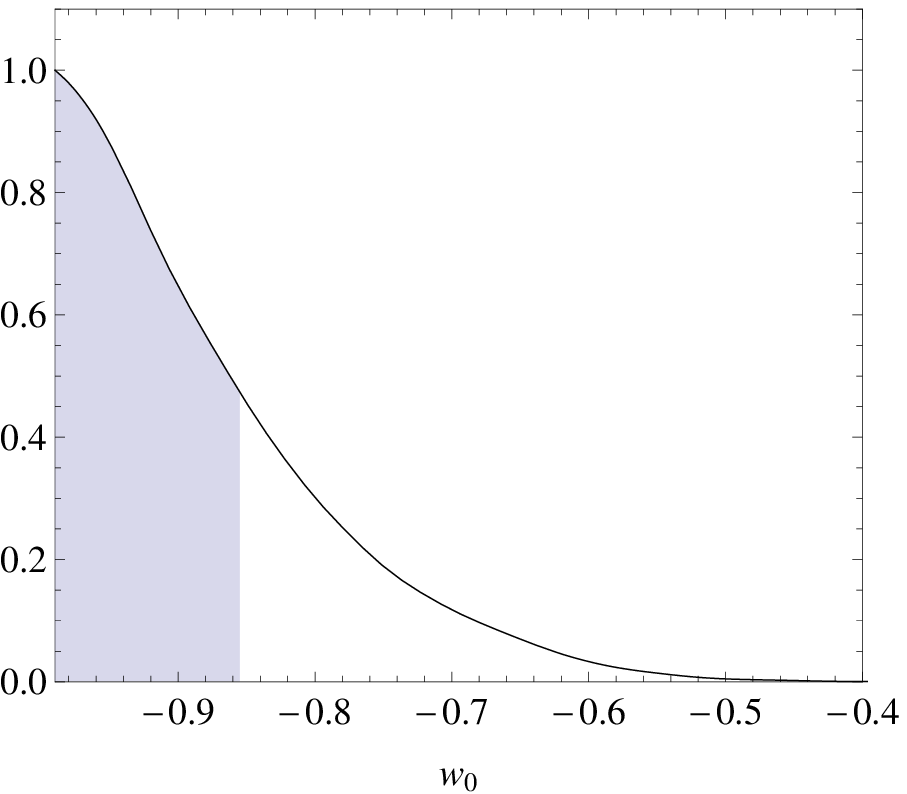}

    \end{minipage}
    \begin{minipage}[t]{3in}
    \centering
    \includegraphics[width=1\columnwidth]{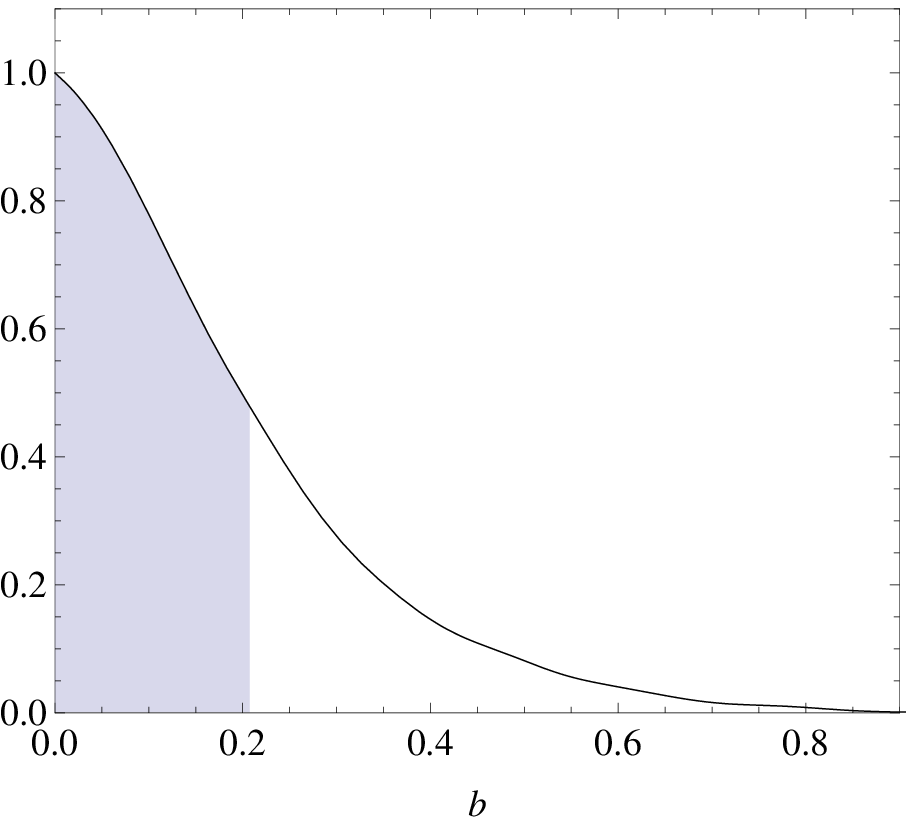}

    \end{minipage}
\end{tabular}
      \begin{tabular}{cc}
    \begin{minipage}[t]{3in}
    \centering
    \includegraphics[width=1\columnwidth]{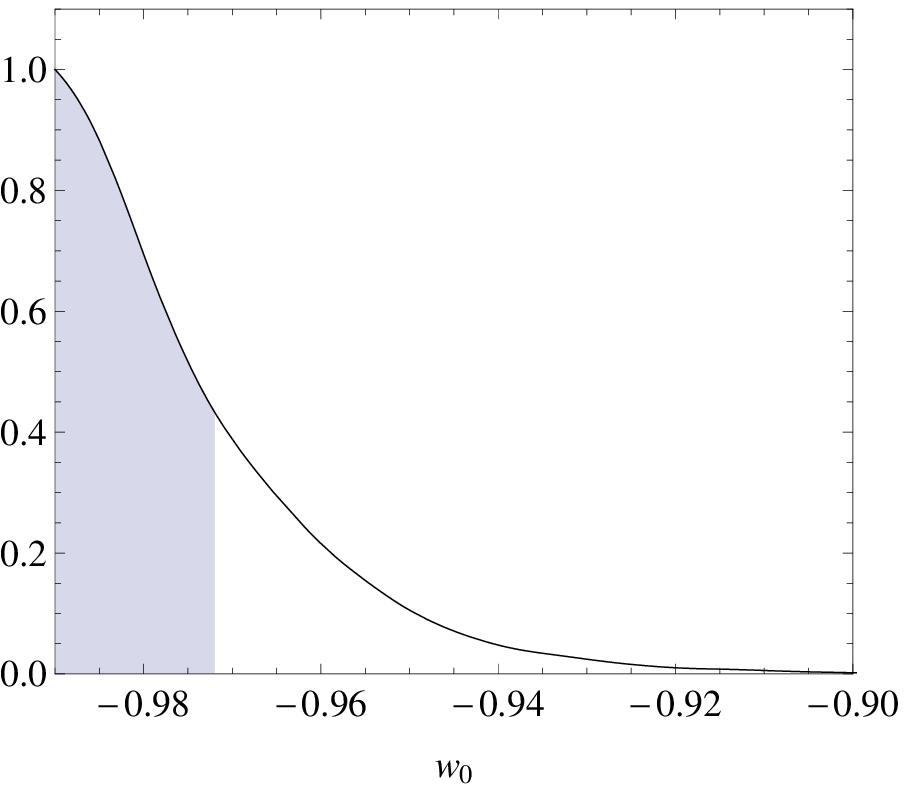}

    \end{minipage}
    \begin{minipage}[t]{3in}
    \centering
    \includegraphics[width=1\columnwidth]{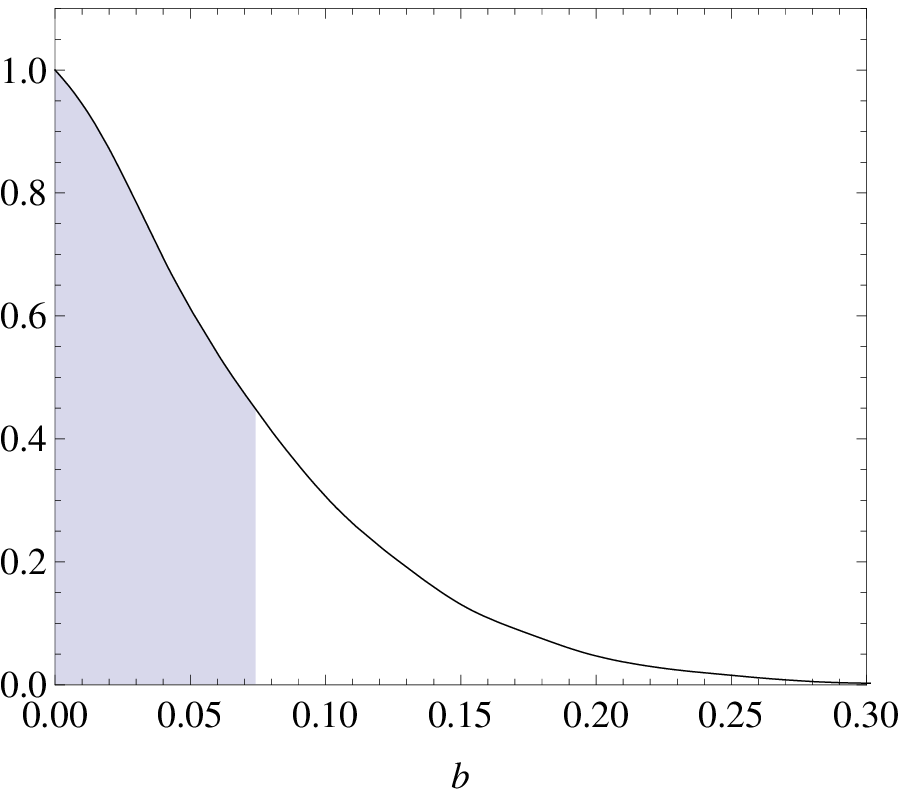}

    \end{minipage}
\end{tabular}
\caption{\label{Model_1_NODE}Fitting results of the
homogeneous EDE model. The upper panel is
from the {\it Planck} data alone, while the lower panel
is from the combined dataset.}
\end{figure}

\begin{figure}[htp]
\begin{tabular}{cc}
    \begin{minipage}[t]{3in}
    \centering
    \includegraphics[width=1\columnwidth]{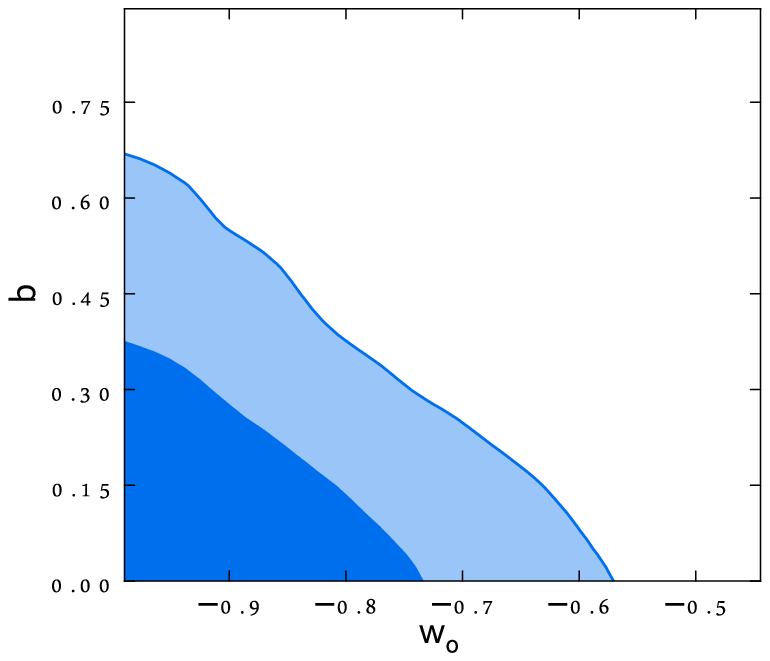}
    \end{minipage}
    \begin{minipage}[t]{3in}
    \centering
    \includegraphics[width=1\columnwidth]{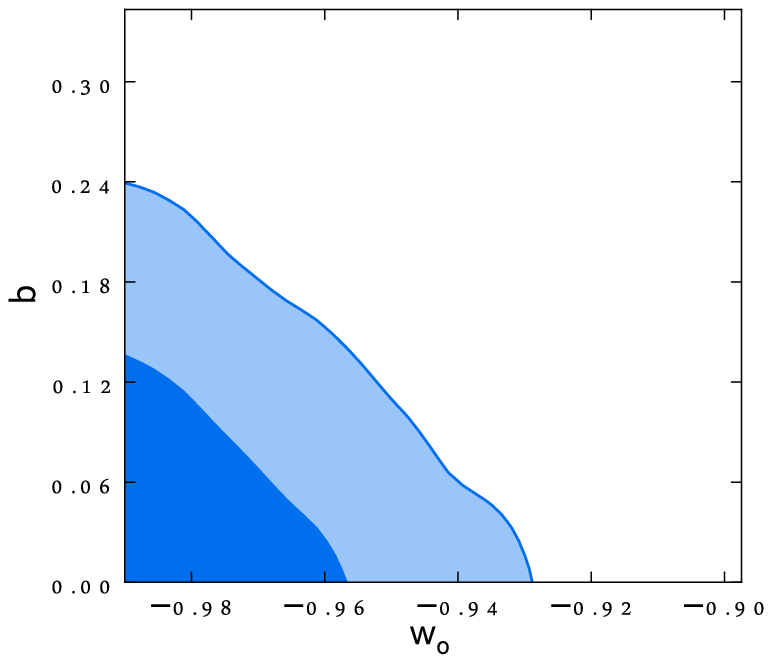}    
    \end{minipage}
\end{tabular}
\caption{\label{2dNP}Fitting results of
the homogeneous EDE model in 2D contour $w_0 - b$. The left panel is
from the {\it Planck} data alone, while the right panel
is from the combined dataset.  }
\end{figure}

\begin{table}[htp]
\centering
\caption[width=1\columnwidth]{\label{Model1_T}Best
fit values and $68\%$ C.L. constraints on the
inhomogeneous EDE.}
\renewcommand\arraystretch{1.0}
\begin{tabular}{p{2cm}<{\centering}p{2cm}<{\centering}p{2cm}<{\centering}p{2cm}<{\centering}p{2cm}<{\centering}}
\hline
\hline
 &\multicolumn{2}{c}{Planck} & \multicolumn{2}{c}{Planck+BAO+SN+H0} \\
\cline{2-5}
 Parameter & Best-fit & $68\%$ limits & Best-fit & $68\%$ limits \\
\hline
$w_0$ & $-0.944$ & $-0.877^{+0.024}_{-0.113}$ & $-0.989$ & $-0.975^{+0.002}_{-0.015}$ \\
$b$ & $0.111$ & $0.171^{+0.033}_{-0.171}$ & $0.012$ & $0.057^{+0.012}_{-0.057}$ \\
\hline
$\chi^2$ & \multicolumn{2}{c}{$9807$} &\multicolumn{2}{c}{$10243$}\\
\hline
\end{tabular}
\end{table}

\begin{table}[htp]
\centering
\caption[width=1\columnwidth]{\label{Model1_T_NODE}Best
fit values and $68\%$ C.L. constraints on the
homogeneous EDE. }
\renewcommand\arraystretch{1.0}
\begin{tabular}{p{2cm}<{\centering}p{2cm}<{\centering}p{2cm}<{\centering}p{2cm}<{\centering}p{2cm}<{\centering}}
\hline
\hline
 &\multicolumn{2}{c}{Planck} & \multicolumn{2}{c}{Planck+BAO+SN+H0} \\
\cline{2-5}
 Parameter & Best-fit & $68\%$ limits & Best-fit & $68\%$ limits \\
\hline
$w_0$ & $-0.886$ & $-0.879^{+0.024}_{-0.111}$ & $-0.988$ & $-0.974^{+0.002}_{-0.016}$ \\
$b$ & $0.010$ & $0.173^{+0.034}_{-0.173}$ & $0.011$ & $0.061^{+0.013}_{-0.061}$ \\
\hline
$\chi^2$ & \multicolumn{2}{c}{$9806$} &\multicolumn{2}{c}{$10243$}\\
\hline
\end{tabular}
\end{table}

\begin{figure}[htp]
      \begin{tabular}{cc}

    \begin{minipage}[t]{2in}
    \centering
    \includegraphics[width=1\columnwidth]{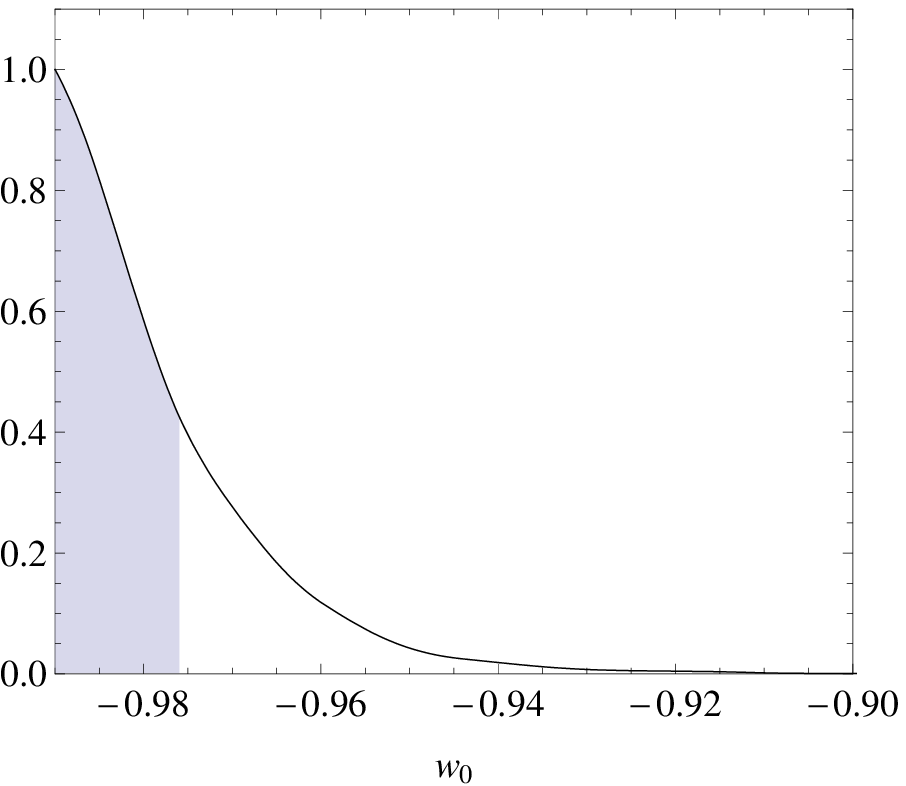}
    \end{minipage}
    \begin{minipage}[t]{2in}
    \centering
    \includegraphics[width=1\columnwidth]{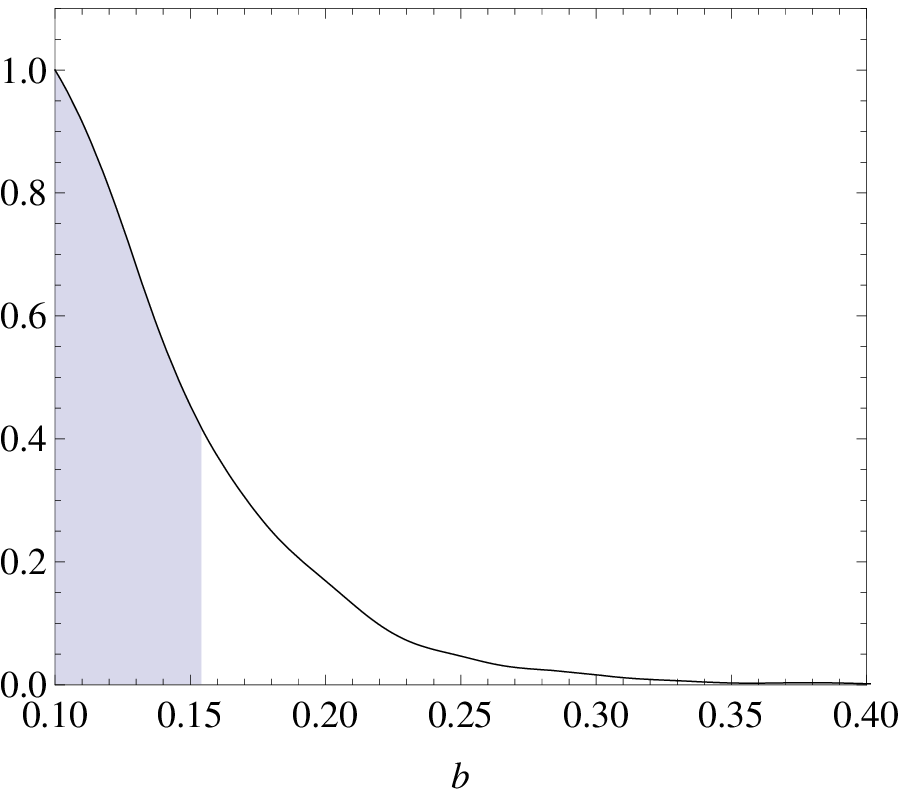}
    \end{minipage}

       \begin{minipage}[t]{2in}
    \centering
    \includegraphics[width=1\columnwidth]{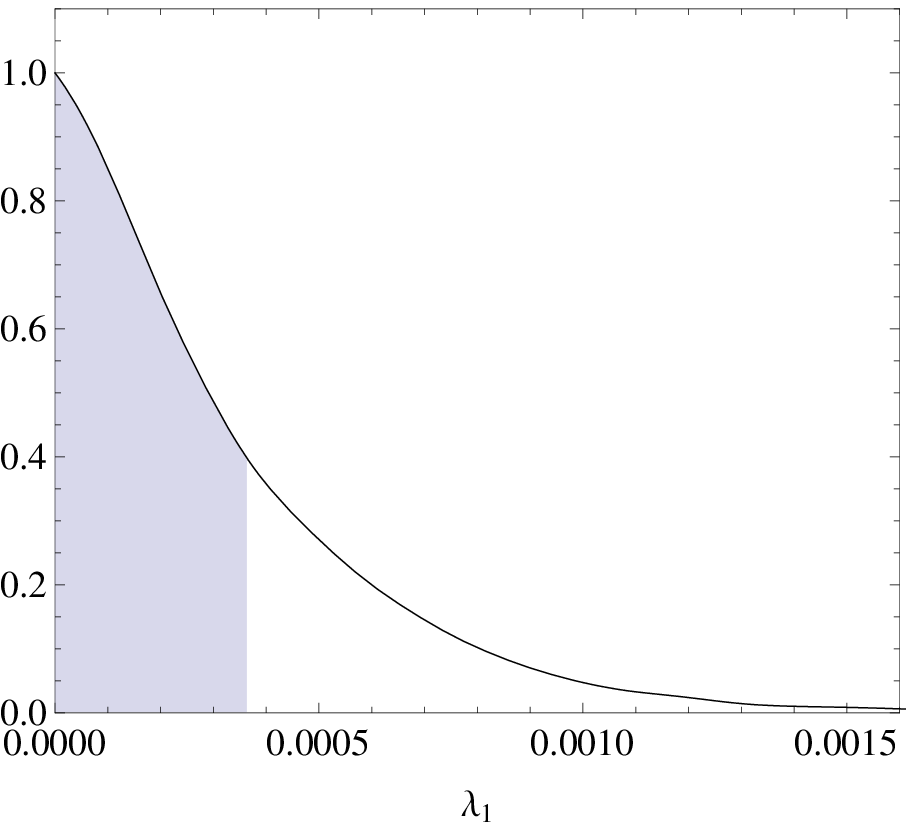}
    \end{minipage}

\end{tabular}
      \begin{tabular}{cc}
    \begin{minipage}[t]{2in}
    \centering
    \includegraphics[width=1\columnwidth]{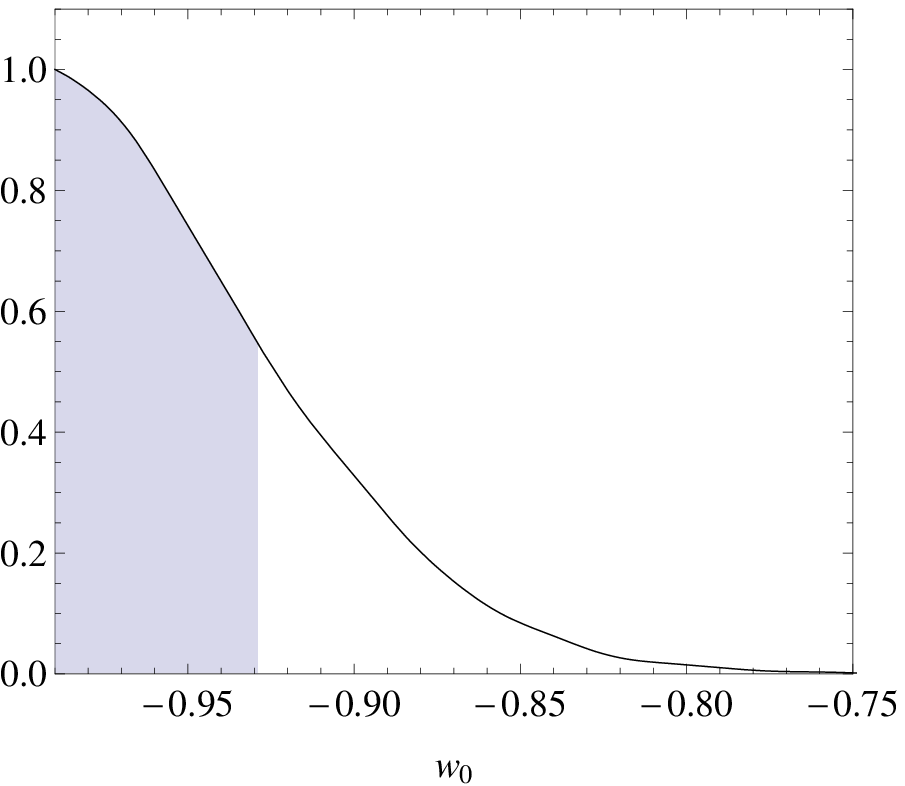}
    \end{minipage}
    \begin{minipage}[t]{2in}
    \centering
    \includegraphics[width=1\columnwidth]{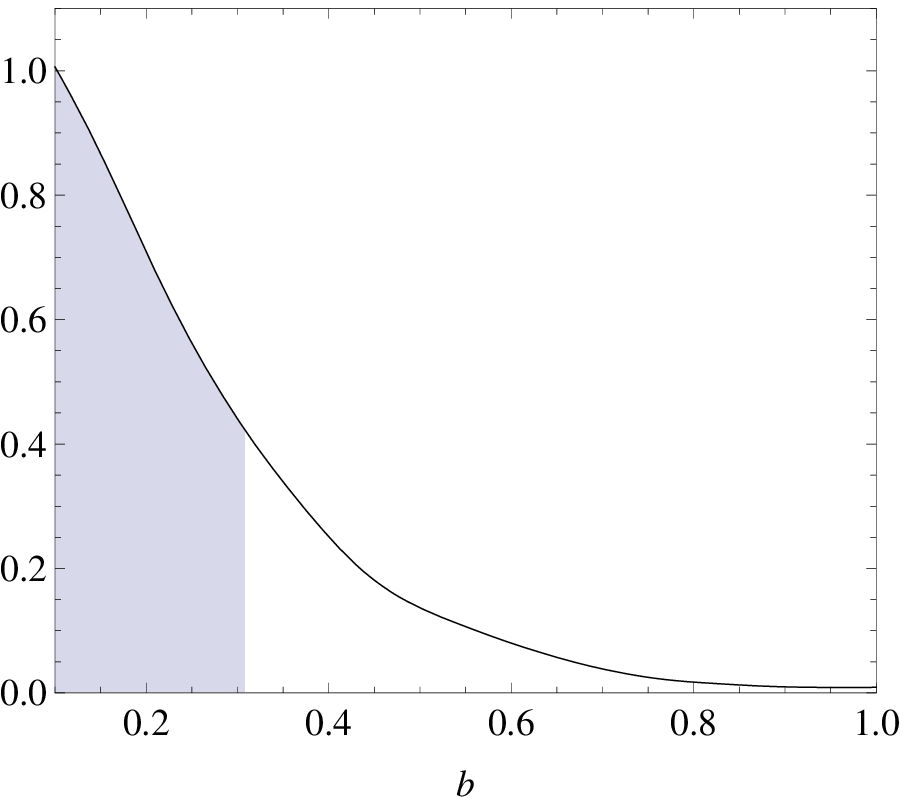}
    \end{minipage}

   \begin{minipage}[t]{2in}
    \centering
    \includegraphics[width=1\columnwidth]{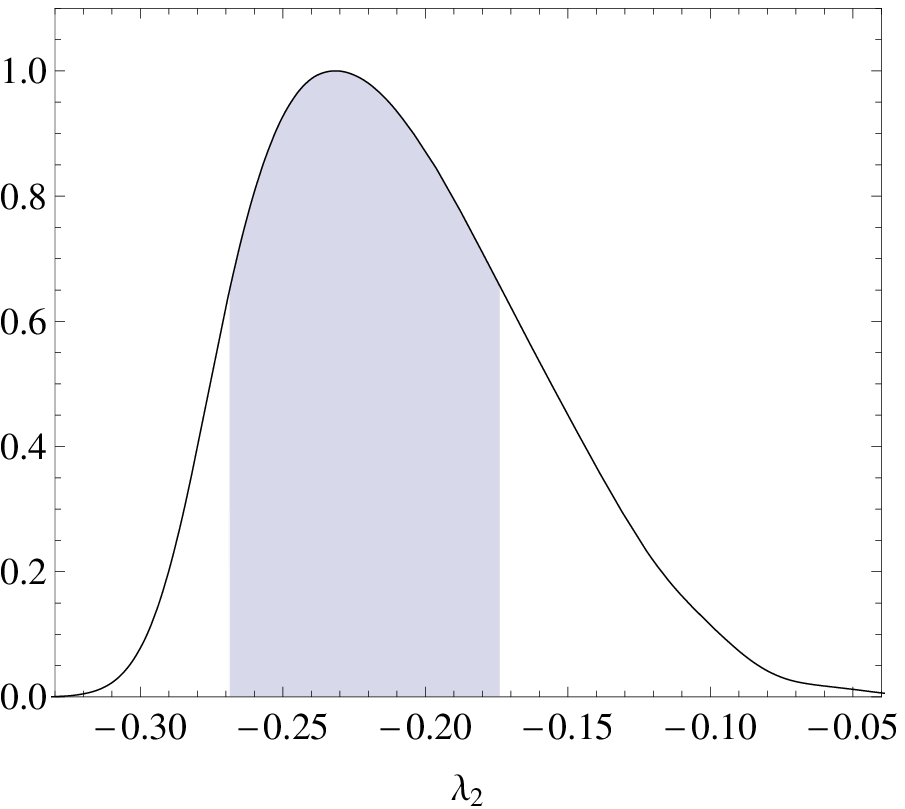}
     \end{minipage}
\end{tabular}
\caption{\label{interaction_fitting}Global
fitting results of the inhomogeneous EDE model with interaction. The upper panel is for the
interaction proportional to the energy density of
DM, while the lower panel is for the interaction
proportional to the energy density of DE.  }

\end{figure}
\begin{figure}[htp]
      \begin{tabular}{cc}

    \begin{minipage}[t]{2in}
    \centering
    \includegraphics[width=1\columnwidth]{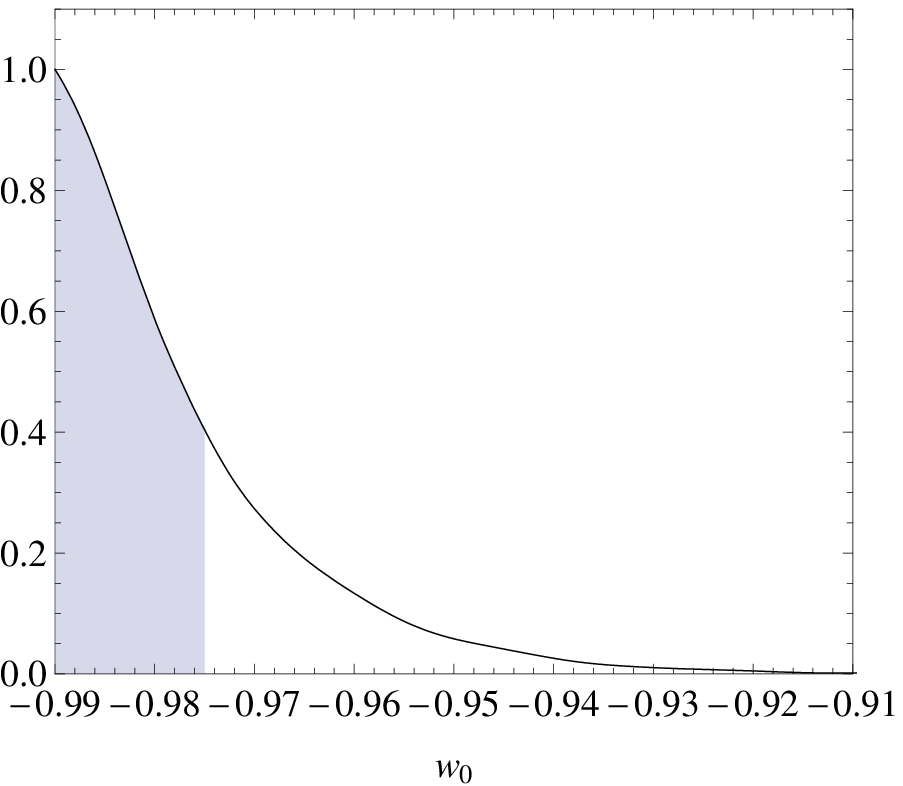}
    \end{minipage}
    \begin{minipage}[t]{2in}
    \centering
    \includegraphics[width=1\columnwidth]{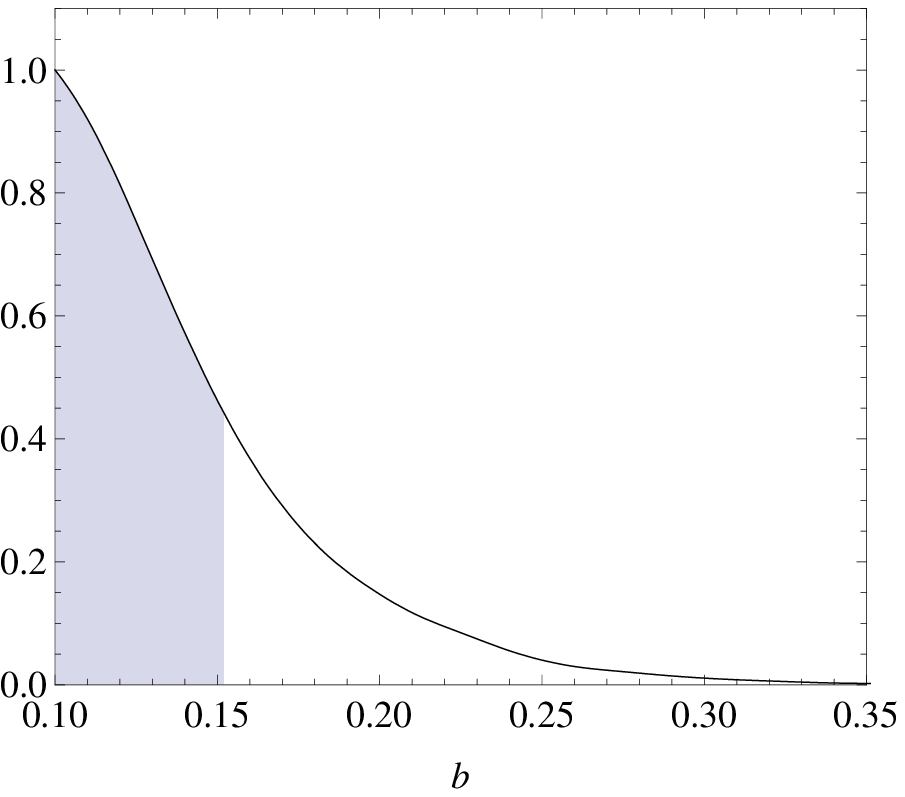}
    \end{minipage}

       \begin{minipage}[t]{2in}
    \centering
    \includegraphics[width=1\columnwidth]{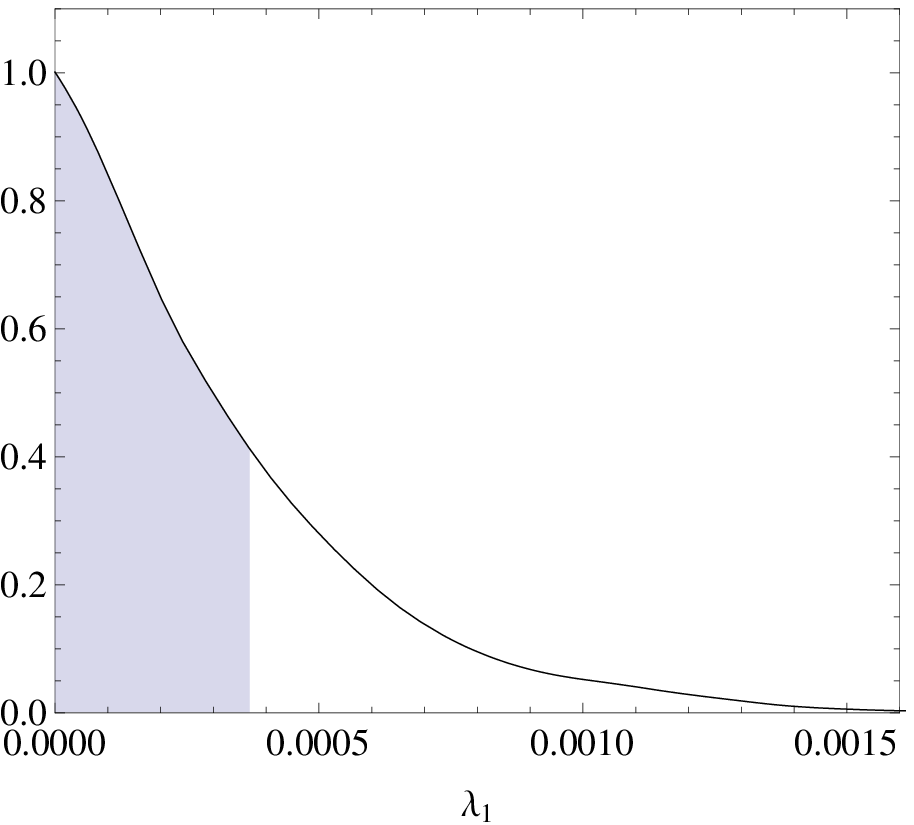}
    \end{minipage}

\end{tabular}
      \begin{tabular}{cc}
    \begin{minipage}[t]{2in}
    \centering
    \includegraphics[width=1\columnwidth]{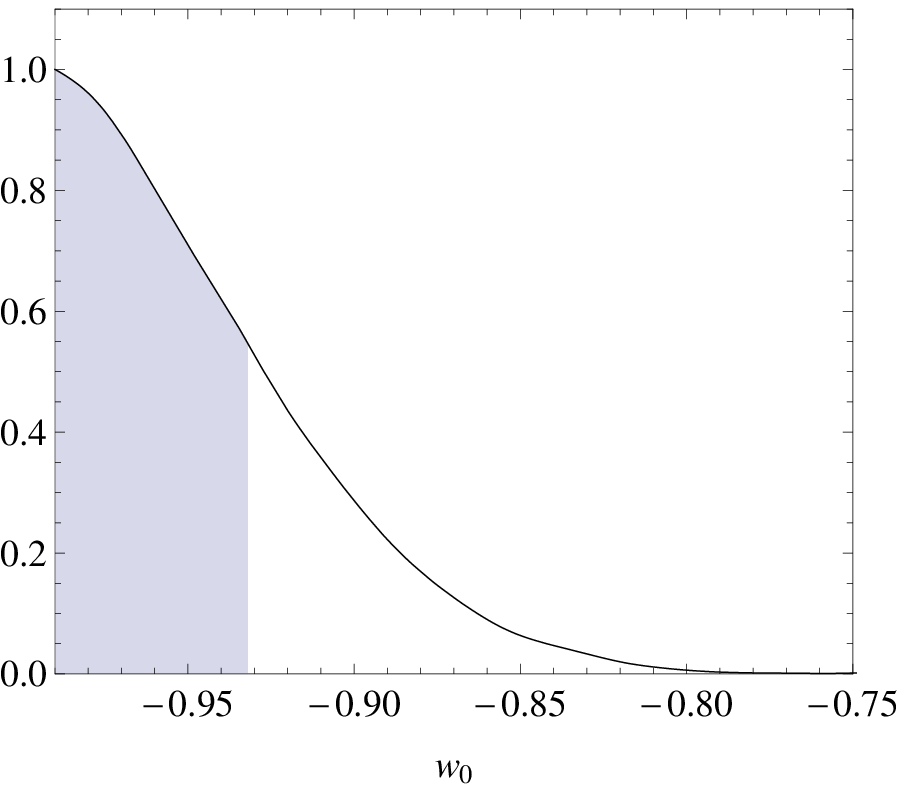}
    \end{minipage}
    \begin{minipage}[t]{2in}
    \centering
    \includegraphics[width=1\columnwidth]{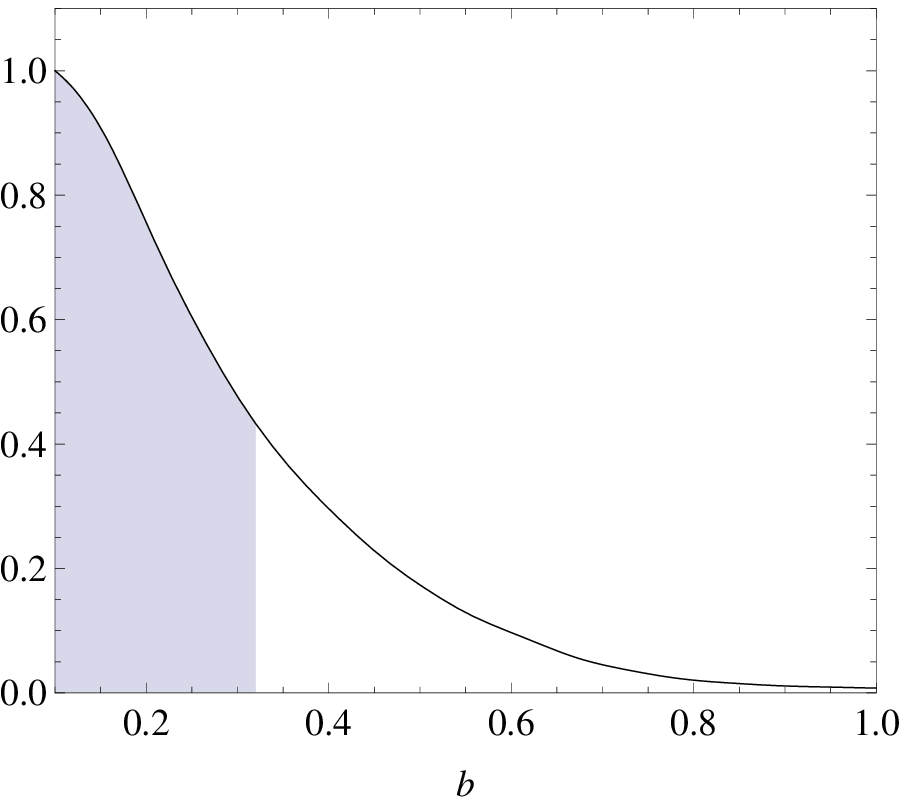}
    \end{minipage}

   \begin{minipage}[t]{2in}
    \centering
    \includegraphics[width=1\columnwidth]{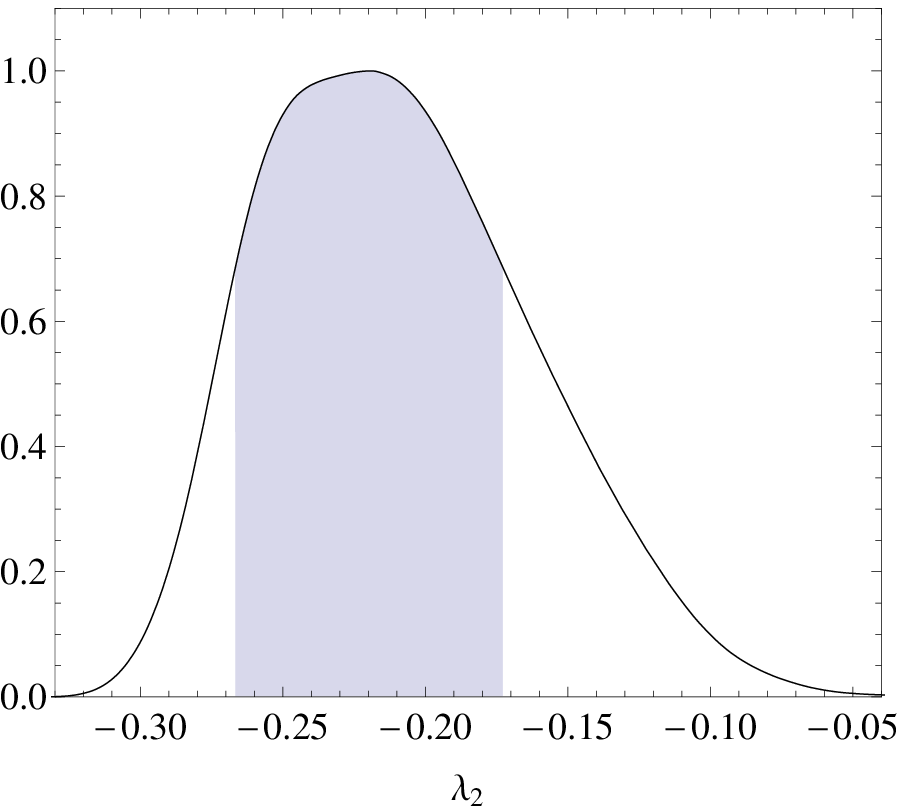}
     \end{minipage}
\end{tabular}
\caption{\label{interaction_fitting_NODE}Global
fitting results of the homogeneous EDE model with interaction. The upper panel is for the interaction
proportional to the energy density of DM, while
the lower panel is for the interaction
proportional to the energy density of DE.  }

\end{figure}

\begin{table}[htp]
\centering
\caption[width=1\columnwidth]{\label{interactions_table}Best
fit values and $68\%$ C.L. constraints on inhomogeneous EDE
models with interaction using the combined dataset of {\it Planck}+BAO+SN+$H_0$}
\renewcommand\arraystretch{1.0}
\begin{tabular}{p{2cm}<{\centering}p{2cm}<{\centering}p{2cm}<{\centering}p{2cm}<{\centering}p{2cm}<{\centering}}
\hline
\hline
 &\multicolumn{2}{c}{Interaction $\propto \rho_{DM}$} & \multicolumn{2}{c}{Interaction $\propto \rho_{DE}$} \\
\cline{2-5}
 Parameter & Best-fit & $68\%$ limits & Best-fit & $68\%$ limits\\
\hline
$w_0$ &  $-0.985$  & $-0.978^{+0.002}_{-0.012}$ &$-0.989$ & $-0.941^{+0.012}_{-0.049}$\\
$b$ & $0.113$ & $0.147^{+0.007}_{-0.047}$ & $0.200$ & $0.274^{+0.034}_{-0.174}$\\
$\lambda_1$ & $0.000274$ & $0.000309^{+0.000054}_{-0.000309}$ &-& -\\
$\lambda_2$ & - & - &$-0.137$& $-0.209^{+0.035}_{-0.050}$  \\
\hline
$\chi^2$ & \multicolumn{2}{c}{$10247$} &\multicolumn{2}{c}{$10238$}\\
\hline
\end{tabular}
\end{table}

\begin{table}[htp]
\centering
\caption[width=1\columnwidth]{\label{interactions_NODE_table}Best
fit values and $68\%$ C.L. constraints on homogeneous
models with interaction using the combined dataset of {\it Planck}+BAO+SN+$H_0$}
\renewcommand\arraystretch{1.0}
\begin{tabular}{p{2cm}<{\centering}p{2cm}<{\centering}p{2cm}<{\centering}p{2cm}<{\centering}p{2cm}<{\centering}}
\hline
\hline
 &\multicolumn{2}{c}{Interaction $\sim \rho_{DM}$} & \multicolumn{2}{c}{Interaction $\sim \rho_{DE}$} \\
\cline{2-5}
 Parameter & Best-fit & $68\%$ limits & Best-fit & $68\%$ limits\\
\hline
$w_0$ &  $-0.974$  & $-0.977^{+0.002}_{-0.013}$ &$-0.966$ & $-0.943^{+0.011}_{-0.047}$\\
$b$ & $0.115$ & $0.144^{+0.008}_{-0.044}$ & $0.123$ & $0.282^{+0.038}_{-0.182}$\\
$\lambda_1$ & $0.000109$ & $0.000311^{+0.000057}_{-0.000311}$ &-& -\\
$\lambda_2$ & - & - &$-0.189$& $-0.209^{+0.036}_{-0.058}$  \\
\hline
$\chi^2$ & \multicolumn{2}{c}{$10248$} &\multicolumn{2}{c}{$10237$}\\
\hline
\end{tabular}
\end{table}

\begin{figure}
\centering
\includegraphics[width=0.5\columnwidth]{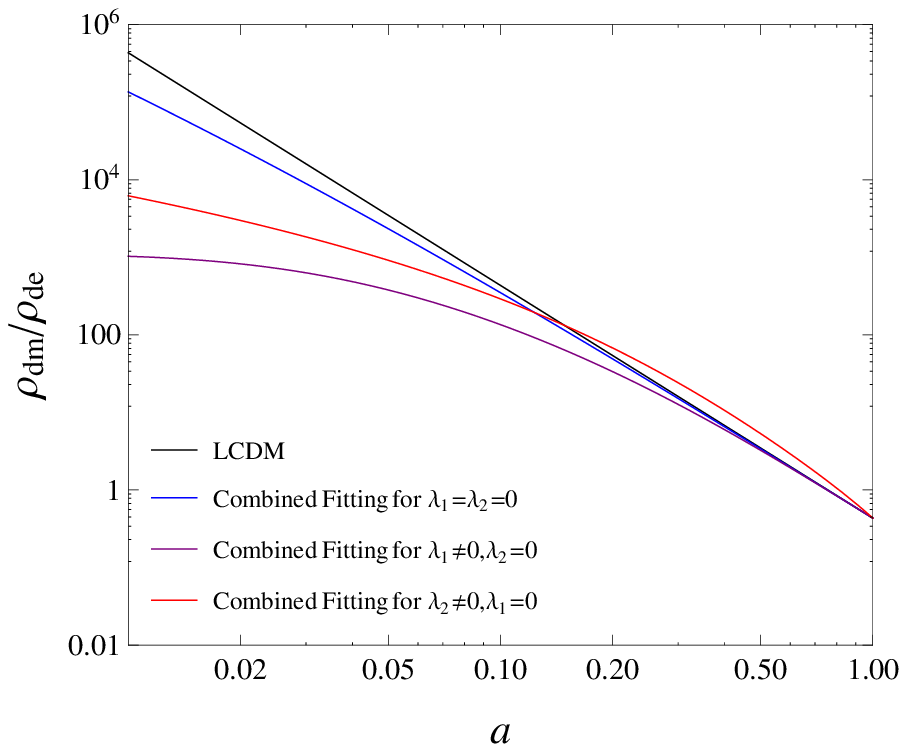}
\caption[width=1\columnwidth]{\label{DM2DE}Ratio of DM energy density to DE energy density of best-fit EDE models.
}
\end{figure}

We first assume that DE and DM evolve
independently in the MCMC analysis. For the
inhomogeneous EDE, with Planck data alone, we show the
results in Table \ref{Model1_T}. The likelihood distribution for
parameters $w_0$ and $b$ in the EDE model are shown
in the upper panel of Figure \ref{Model_1}. Using the combined
dataset, we can see how the constraints improve.
We list the results in Table \ref{Model1_T} and exhibit the
likelihood distributions of $w_0$ and $b$ in the lower panel of Figure \ref{Model_1}. It
is easy to see that the addition of the complementary
data clearly improves the constraints on the
EDE parameters. This is because the parameters
which could be degenerate with the EDE
parameters, such as the Hubble parameter, are
well-constrained by other observations. The 2D contour for $w_0$-$b$ is shown in Figure \ref{2d}.

We then turn to the case where DE perturbations
are neglected. Performing an analysis with Planck
data alone, we show the fitting results in Table
\ref{Model1_T_NODE} and likelihoods for the EDE
model parameters in the upper panel of Figure
\ref{Model_1_NODE}. We find that with Planck
data alone, the best-fit of $w_0$ is farther away from $-1$ and the best-fit of $b$ gets smaller than the results of the
inhomogeneous case. But the mean values and $68\%$ limits are nearly the same. With combined
data sets, the best-fit and $68\%$ limit of the inhomogeneous and homogeneous cases shows basically no difference.
The 2D contour of $w_0$ - $b$ is shown in Figure \ref{2dNP}.
Both of them suggest that $w_0$ is very close to $-1$ and $b$ is small which inmplies tiny EDE effect.

Considering the interaction between DE and DM, we
carry out the MCMC analysis again. We display the
likelihood distributions of the EDE parameters and the coupling strength from the
combined datasets for inhomogeneous and
homogeneous EDE in Figure \ref{interaction_fitting}
and \ref{interaction_fitting_NODE}, respectively. The best fitting
values and $68\%$ limits are listed in Table
\ref{interactions_table} and \ref{interactions_NODE_table}. For inhomogeneous EDE with different forms of interaction,
the fitting results of $w_0$
are similar to that of non-interacting EDE model. On the other hand, we see that in the presence of interaction, the limit of $b$ is larger, which implies that the EDE effect is stronger. This may be attributed to the choice of prior $b$(II). But this tendency is also clear when the interaction is proportional to DE density, which share the same prior of $b$ as non-interacting EDE models.

In the theoretical discussion of the CMB spectrum, we see that
the interaction proportional to DM energy density has stronger impact on CMB power spectrum than the interaction proportional to
DE energy density. This can be seen also in the fitting results as we find that the limit of $\lambda_1$ is
much smaller than that of $\lambda_2$. The negative best-fit value of $\lambda_2$ agrees with the result when DE EoS is constant \cite{1311.7380}. The fitting results for the homogeneous case are similar to the inhomogeneous case.

In the tables of the fitting results, we presented the $\chi^2$ for the best-fit models. When the interaction is proportional to the energy density of DM, the prior of $b$ and $\lambda_1$ is highly limited as mentioned above. As a result, the $\chi^2$ is a bit larger than the non-interacting EDE model. On the contrary, the presence of the interaction proportional to DE density decreases the $\chi^2$. Comparing with $\Lambda$CDM model, in which $\chi^2=9806, 10242$ for $Planck$ and the combined dataset of $Planck$+BAO+SN+$H_0$ respectively, we find that the EDE models and its interactions with DM are compatible with current observations.

To examine whether the EDE models allowed by the observations is effective to alleviate the coincidence
problem, we plot in Figure \ref{DM2DE} the ratio
of DM energy density to DE energy density in the best-fit EDE models of the joint analysis and compare them
with the $\Lambda$CDM prediction.
Comparing the EDE models with $\Lambda$CDM model, we
find that if the interaction is proportional to the energy
density of DM, the ratio evolves slower than those in other models. By introducing the interaction between dark sectors, the coincidence problem becomes less acute. We conclude that the EDE model is
compatible with observations and it is effective to
alleviate the coincidence problem.

\section{conclusions and discussions}

In this paper we have studied the influence of
EDE on DM perturbations. We have observed that,
different from DE models with constant EoS, DM perturbation is larger in inhomogeneous EDE models than in homogeneous
EDE model in which DE fluctuation is neglected. We have also
disclosed the difference between inhomogeneous and
homogeneous EDE in the large scale CMB power
spectrum. It is expected that the probe of the growth of large scale structure
and small $l$ CMB power spectrum can help to
distinguish homogeneous and inhomogeneous
EDE models.

Furthermore we have extended our discussion to
the interaction between EDE and DM. We
have observed that an interaction between EDE and DM
also affects DM perturbations and small $l$
CMB power spectrum, which may be degenerate with the
effect of DE fluctuations. Comparing these
effects, we found that the interaction
between EDE and DM has stronger influence on DM
perturbations and on the ISW effect.

We have confronted the EDE models to
{\it Planck} data and a combined dataset of {\it Planck}+BAO+SN+$H_0$. The analysis showed
that the coincidence problem in all best-fit EDE models
is less severe than in $\Lambda$CDM model. The
positive coupling between EDE and DM proportional
to the energy density of DM is particularly effective to alleviate the coincidence problem. This can be clearly seen in Fig. 18 that with the positive coupling between EDE and DM proportional
to the energy density of DM, it has longer period for the DE and DM to be comparable.

It is interesting to further examine whether the
disclosed impacts of the DE fluctuations and the
interaction between DE and DM on observables are
specific to EDE models. A lot of efforts on this
problem are called for.

\begin{acknowledgments}
We acknowledge financial supports from National
Basic Research Program of China (973 Program
2013CB834900) and National Natural Science
Foundation of China. E.A. acknowledges financial
support from CNPq (Conselho Nacional de
Desenvolvimento Cient\'\i fico e Tecnol\' ogico)
and from FAPESP (Funda\c c\~ao de Amparo \`a
Pesquisa do Estado de S\~ao Paulo).

\end{acknowledgments}


\begin{thebibliography}{}

\bibitem{coincidence}S. Weinberg, Rev. Mod. Phys. 61 1(1989).
\bibitem{coin} L. P. Chimento, A. S. Jakubi, D. Pavon, and W. Zimdahl, Phys. Rev. D67, 083513 (2003),
arXiv: astro-ph/0303145 [astro-ph].
\bibitem{1304.3724}A. Hojjati, E. V. Linder, J. Samsing, Phys.Rev. Lett111, 041301(2013), arXiv: 1304.3724
\bibitem{Doran2001} M. Doran,  M. Lilley, J. Schwindt, C. Wetterich, ApJ. 559 501(2001)
\bibitem{1003.1259} U. Alam, ApJ. 714 1460, arXiv: 1003.1259(2010)
\bibitem{astro-ph/0508132}M. Doran, K. Karwan, C. Wetterich, JCAP. 0511 007(2005), arXiv: astro-ph/050813
\bibitem{astro-ph/0611507}P. Wu, H. Yu, Phys. Lett. B643 315(2006), arXiv: astro-ph/0611507
\bibitem{1305.2209}J. Bielefeld, W.L.Kimmy Wu, R.R. Caldwell, O. Dore, Phys.Rev. D88 103004(2013), arXiv: 1305.2209
\bibitem{0901.0605}J.Q Xia, M. Viel, JCAP. 0904 002(2009), arXiv: 0901.0605
\bibitem{Muller}C.M. M\"uller, G. Sch\"afer, C. Wetterich, Phys.Rev. D70 083504(2004)
\bibitem{Bartelmann}M. Bartelmann, M. Doran, C. Wetterich, A\&A, 454 27(2006)
\bibitem{1004.0437}U. Alam, Z. Lukic, S. Bhattacharya, ApJ. 727 87(2011)
\bibitem{1207.1723}F. Fontanot, V. Springel, R. E. Angulo, B. Henriques, MNRAS. 426 2335(2012)
\bibitem{0809.3404}M. Grossi, V. Springel, MNRAS. 394 1559(2009)
\bibitem{Rahvar}M. Sadegh Movahed, S. Rahvar, Phys. Rev. D73, 083518 (2006)
\bibitem{yungui} Y. G. Gong, Class. Quan. Grav. 22, 2121 (2005)
\bibitem{1303.0414}R. C. Batista, F. Pace, JCAP. 1306 044(2013)
\bibitem{xx} S. Das, A. Shafieloo, T. Souradeep, JCAP, 10, 016 (2013), arXiv: 1305.4530
\bibitem{Baldi}M. Baldi, MNRAS. 422 1028(2012)
\bibitem{wetterich}C. Wetterich, Phys.Lett. B594 17(2004)
\bibitem{1311.7380}A.A. Costa, X.D. Xu, B. Wang, E.G.M. Ferreira, E. Abdalla, Phys.Rev. D89 103531(2014)
\bibitem{1308.1475}X.D. Xu, B. Wang, P.J. Zhang, F. Atrio-Barandela, JCAP, 1312 001(2013), arXiv: 1308.1475
\bibitem{1012.3904}J.H. He, B. Wang, E. Abdalla, Phys.Rev. D83 063515(2011), arXiv: 1012.3904
\bibitem{1001.0097}J.H. He, B. Wang, E. Abdalla, D. Pavon, JCAP. 1012 022(2010), arXiv: 1001.0079
\bibitem{0710.1198}E. Abdalla, L. R. W. Abramo, J. Sodre, L., B. Wang, Phys.Lett. B673 107(2009), arXiv: 0710.1198
\bibitem{abdferreirawang}E. Abdalla, E.G. M. Ferreira, J. Quintin and Bin Wang, arXiv: 1412.2777
\bibitem{1202.0499}E. Abdalla, L. Graef, B. Wang, Phys.Lett. B726 786(2012), arXiv: 1202.0499
\bibitem{Sandro1}S. Micheletti, E. Abdalla, B Wang, Phys.Rev. D79 123506(2009)
\bibitem{Sandro2}S. Micheletti, JCAP. 1005 009(2010)
\bibitem{1206.2589}O. Bertolami, P. Carrilho, J. Paramos, Phys.Rev. D86 103522(2012)
\bibitem{0807.3471}J.H. He, B. Wang, E. Abdalla, Phys.Lett. B671 139(2009), arXiv: 0807.3471
\bibitem{0902.0660}J.H. He, B. Wang, Y.P. Jing, JCAP. 0907 030(2009), arXiv: 0902.0660
\bibitem{0906.0677}J.H. He, B. Wang, P.J. Zhang, Phys.Rev. D80 063530(2009), arXiv: 0906.0677
\bibitem{camb}A. Lewis, A. Challinor, A. Lasenby, Astrophys.J. 538, 473(2000), arXiv: astro-ph/9911177 [astro-ph]
\bibitem{cosmomc1}A. Lewis, S. Bridle, Phys.Rev. D66 103511(2002), arXiv: astro-ph/0205436 [astro-ph]
\bibitem{cosmomc2}A. Lewis, Phys.Rev. D87 103529(2013), arXiv: 1304.4473 [astro-ph.CO]
\bibitem{planck1}Planck Collaboration I. A\&A 571, A1(2014), arXiv: 1303.5062
\bibitem{planck2}Planck Collaboration XVI. A\&A 571, A16(2014), arXiv: 1303.5076
\bibitem{planck3}Planck Collaboration XV. A\&A 571, A15(2014), arXiv: 1303.5075
\bibitem{bao1}F. Beutler, C. Blake, M. Colless, D. H. Jones, L. Staveley-Smith, et al., MNRAS. 416 3017(2011), arXiv: 1106.3366
\bibitem{bao2}N. Padmanabhan, X. Xu, D. J. Eisenstein, R. Scalzo, A. J. Cuesta, et al., MNRAS. 427 2132(2012), arXiv: 1202.0090
\bibitem{bao3} L. Anderson, E. Aubourg, S. Bailey, D. Bizyaev, M. Blanton, et al., MNRAS. 427 3435(2013), arXiv: 1203.6594
\bibitem{SN}N. Suzuki, D. Rubin, C. Lidman, G. Aldering, R. Amanullah, et al., Apj. 746 85(2012), arXiv: 1105.3470
\bibitem{H0}A. G. Riess, L. Macri, S. Casertano, H. Lampeitl, H. C. Ferguson, et al., Apj. 730 119(2011), arXiv: 1103.2976
\bibitem{Xu2011}X.D. Xu, J.H. He and B. Wang, Phys.Lett B 701 (2011) 513¨C519, arXiv: 1103.2632


\end{thebibliography}
\end{document}